\documentclass[letterpaper]{article}

\usepackage[T1]{fontenc}

\usepackage{geometry}
\usepackage{setspace}

\usepackage[
  backend=biber,
  style=chem-acs,
  articletitle=true,
  giveninits=true,
  doi=true,
  url=false,
  isbn=false,
  eprint=false
]{biblatex}
\AtEveryBibitem{\clearfield{note}}
\usepackage{graphicx}
\usepackage{float}
\newfloat{scheme}{htbp}{los}
\floatname{scheme}{Scheme}
\floatname{chart}{Chart}
\newfloat{graph}{htbp}{loh}

\usepackage{chemformula} 
\usepackage[version = 4]{mhchem} 

\newcommand{\xGNR}{(4,2,7)-chGNR}

\usepackage{siunitx}    
\DeclareSIUnit\angstrom{\text{\AA}}
\DeclareSIUnit\bar{bar}

\usepackage{epstopdf} 

\usepackage{pdfpages}
\usepackage{authblk}
\author[$\dagger$,1,2]{Xuanchen Li}
\author[$\dagger$,1]{Amogh Kinikar}
\author[$\dagger$,3]{Vikas Sharma}
\author[1]{Andres Ortega-Guerrero}
\author[1,4]{Riya Sebait}
\author[3]{George F. S. Whitehead}
\author[2,4,5]{Mickael Lucien Perrin}
\author[1]{Carlo A. Pignedoli}
\author[1,6]{Roman Fasel}
\author[$*$,3,7,8]{Ashok Keerthi}
\author[$*$,1]{Gabriela Borin Barin}

\affil[1]{nanotech@surfaces Laboratory, Empa — Swiss Federal Laboratories for Materials Science and Technology, Dübendorf, Switzerland}
\affil[2]{Department of Information Technology and Electrical Engineering, ETH Zürich, Zürich, Switzerland}
\affil[3]{Department of Chemistry, School of Natural Sciences, The University of Manchester, Manchester, UK}
\affil[4]{Transport at Nanoscale Interfaces Laboratory, Empa — Swiss Federal Laboratories for Materials Science and Technology, Dübendorf, Switzerland}
\affil[5]{Quantum Center, ETH Zürich, Zürich, Switzerland}
\affil[6]{Department of Chemistry, Biochemistry and Pharmaceutical Sciences, University of Bern, Bern, Switzerland}
\affil[7]{Photon Science Institute, The University of Manchester, Manchester, UK}
\affil[8]{National Graphene Institute, The University of Manchester, Manchester, UK}
\affil[$\dagger$]{These authors contributed equally.}

\title{Electronic and Vibrational Properties of On‑Surface Synthesized Gulf‑Edged Chiral Graphene Nanoribbons}
\date{*Email: gabriela.borin-barin@empa.ch, ashok.keerthi@manchester.ac.uk}

\begin{document}

\maketitle

\begin{abstract}
  On-surface synthesis enables graphene nanoribbons (GNRs) with atomic precision, but the structural diversity of chiral GNRs remains constrained by the limited range of precursor architectures. Here, we design a trisnaphthalene-based diiodo precursor and use it to synthesize a gulf-edged (4,2,7)-chGNR on Au(111). Scanning tunneling microscopy and bond-resolved non-contact atomic force microscopy establish the atomically precise ribbon structure, while scanning tunneling spectroscopy and periodic density functional theory calculations identify a closed-shell semiconducting state with an experimental band gap of \SI{1.8}{\electronvolt}. Raman spectroscopy, supported by vibrational calculations, resolves a mode localized predominantly at the gulf-edge C--H groups and reveals rapid spectral degradation following air exposure. These results demonstrate a precursor-design concept for a distinct chGNR architecture and correlate its atomic structure with its electronic, vibrational, and environmental-response properties.
\end{abstract}

\begin{center}
\end{center}

\section{Introduction}

Graphene nanoribbons (GNRs) are narrow strips of graphene with nanometer-scale widths. Unlike pristine graphene, which is gapless, GNRs typically exhibit a band gap due to quantum confinement, and their electronic, optical, and magnetic properties are highly tunable through structural modifications~\cite{yano_quest_2020}. This degree of tunability within a pure carbon framework makes GNRs promising candidates for next-generation electronic~\cite{radsar_graphene_2021}, thermoelectric~\cite {zheng_enhanced_2012}, spintronic~\cite{zhang_spin_2021,guo_field_2008}, and optical~\cite {kumar_electronic_2023} devices.

Various types of GNRs have been established based on their edge topology. Armchair GNRs (AGNRs) feature pure armchair edges and behave as semiconductors with width-dependent band gaps. They are classified into three families, 3p, 3p + 1, and 3p + 2, based on the number of carbon atoms across their width, where p is an integer. Within each family, the band gap decreases as the ribbon width increases~\cite {yang_quasiparticle_2007,deniz_revealing_2017}. Zigzag GNRs (ZGNRs), characterized by zigzag edges, exhibit spin-polarized edge states that couple ferromagnetically along each edge and antiferromagnetically between opposite edges~\cite{brede_detecting_2023}. Chiral GNRs (chGNRs) combine armchair and zigzag segments, offering additional flexibility for tuning band gaps and carrier mobility~\cite{liu_cove-edged_2024} and serve as platforms for exploring phenomena such as topological phase transitions~\cite{li_topological_2021} and magnetic exchange coupling~\cite{wang_magnetic_2022}. Furthermore, more complex edge structures, including cove~\cite{liu_toward_2015}, gulf~\cite{narita_synthesis_2014}, fjord~\cite{yao_synthesis_2021}, chevron~\cite{cai_atomically_2010}, and mixed-edge configurations~\cite{niu_curved_2020,keerthi_-surface_2020}, have also been reported. Beyond these structures, heteroatom-doped GNRs~\cite{gao_heteroatom-doped_2024}, GNR heterojunctions~\cite{cai_graphene_2014}, and porous GNRs~\cite{fan_bottom-up_2024} provide further opportunities to tailor functional properties through structural and chemical modifications~\cite{yin_-surface_2023}.

Realizing these structures with atomic precision requires a bottom-up approach. On-surface synthesis fulfills this requirement by depositing rationally designed molecular precursors onto a catalytic metal substrate under ultra-high vacuum (UHV) conditions, followed by annealing-induced polymerization and cyclodehydrogenation~\cite{cai_atomically_2010}. Beyond delivering atomic precision, this method is directly compatible with UHV high-resolution characterization techniques, including scanning tunneling microscopy and spectroscopy (STM/STS), and non-contact atomic force microscopy (nc-AFM)~\cite{gross_chemical_2009}, enabling in-situ access to both the structure and electronic properties of the resulting GNRs~\cite{koenhoutsma_atomically_2021}.
For device integration, however, GNRs must typically be transferred from UHV to ambient conditions and non-metallic substrates~\cite{chen_graphene_2020}. Raman spectroscopy is routinely employed to monitor their structural integrity and ambient stability under such conditions, as it probes highly structure-sensitive vibrational modes and remains compatible across a wide range of substrates and environmental conditions~\cite{verzhbitskiy_raman_2016,borin_barin_surface-synthesized_2019}.

Alongside the progress in GNR research, multiple on-surface synthesis motifs of the chGNRs have been reported~\cite{keerthi_-surface_2020,deniz_electronic_2025,han_bottom-up_2014,li_topological_2021,wang_bottom-up_2018,de_oteyza_substrate-independent_2016}. However, most chGNR precursors have been designed as halogen-functionalized, $\pi$-extended acene oligomers in which multiple acene units are laterally fused in a regular fashion~\cite{li_topological_2021,deniz_electronic_2025,han_bottom-up_2014,de_oteyza_substrate-independent_2016}. Such designs limit the formation of chGNRs with more complex edge structures, such as mixed edges~\cite{keerthi_-surface_2020} and cove edges~\cite{liu_cove-edged_2024}. Achieving gulf-edged chGNRs requires an alternative synthetic motif.

Nevertheless, the number of molecular precursor architectures capable of encoding periodically indented edge structures in chGNRs remains limited. Here, we employ a trisnaphthalene-based precursor bearing biphenyl substituents and terminal iodine groups to access the gulf-edged (4,2,7)-chGNR on Au(111). The strategic positioning of iodo-groups on the biphenyl moiety ensures a smooth polymerization reaction during on-surface synthesis, while the expanded $\pi$-conjugated area enhances surface anchoring and thereby minimizes desorption and decomposition up to \SI{673}{\kelvin}. This rational design enables the direct synthesis of a well-defined \xGNR{}, with controlled regiospecificity, width uniformity, and edge purity, and establishes a new precursor design methodology for chGNRs more generally. 

The on-surface synthesis of \xGNR{} was characterized by STM, which monitors the polymerization and dehydrogenation steps, and by nc-AFM, which confirms the atomically precise structure. STS measurements combined with density functional theory (DFT) were used to resolve the \xGNR{} band structure, while Raman spectroscopy probed the vibrational modes and ambient stability of the ribbon. Despite its relatively large band gap and non-spin-polarized edges, the \xGNR{} exhibits poor ambient stability, consistent with previous findings that zigzag edge segments can drive ambient instability even in GNRs with predominantly closed-shell electronic structure~\cite{berdonces-layunta_chemical_2021}. 

\section{Results and Discussion}

\subsection{Synthesis of molecular precursor and surface-assisted growth of \xGNR{}}

\begin{figure}[h]
    \centering
    \includegraphics[width=1\linewidth]{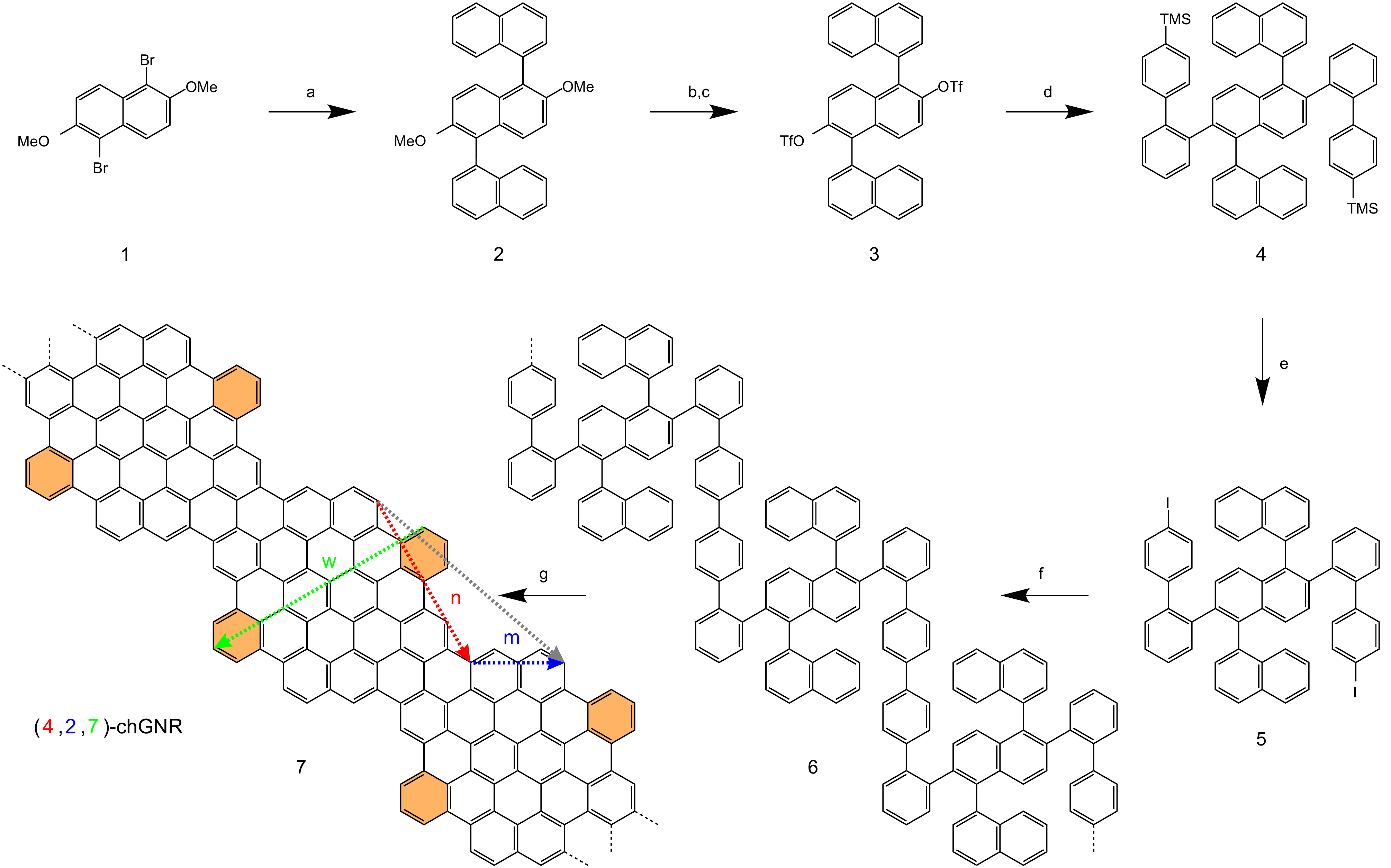}
    \caption{Synthetic route towards the \xGNR{}. Reagents and
    conditions: 
    \textbf{(a)} 4,4,5,5-tetramethyl-2-(naphthalen-1-yl)-1,3,2-dioxaborolane, K$_2$CO$_3$, Pd$_2$(dba)$_3$, SPhos, toluene, water, ethanol, \SI{383}{\kelvin} , 18h, 80\%. 
    \textbf{(b)} BBr$_3$, DCM, \SI{273}{\kelvin}  to \SI{295}{\kelvin} , 18h, ~99\%. 
    \textbf{(c)} trifluoromethanesulfonic anhydride, pyridine, DCM, rt, 12h, 72\%. 
    \textbf{(d)} Compound \textbf{1c}, K$_3$PO$_4$, dry toluene, SPhos, Pd(dppf)Cl$_2$, \SI{373}{\kelvin} , 24h, 80\% (details on \textbf{1c} are provided in the Supporting Information). 
    \textbf{(e)} ICl, DCM, \SI{203}{\kelvin}, 2h, 93\%. 
    \textbf{(f)} Sublimation onto Au(111) , then annealing at \SI{438}{\kelvin} for 30 min. 
    \textbf{(g)} Annealing at \SI{673}{\kelvin} for 30 min. The structure of \xGNR{} is shown and labeled as \textbf{7}. The chiral vector $\mathbf{C}_{\mathrm{h}}(n,m)$ and the width $w$ are indicated with arrows. The benzene ring responsible for forming the gulf edge is highlighted with color. A detailed nomenclature of chGNRs is provided in the Supporting Information.
    }
    \label{fig:reaction_scheme}
\end{figure}

The synthetic route to \xGNR{} and its precursor molecules is illustrated in Figure~\ref{fig:reaction_scheme}. 
Compound (\textbf{2}) (2',6'-dimethoxy-1,1':5',1''-ternaphthalene) was obtained in excellent yield (80\%) from a twofold Suzuki coupling of naphthalene boronic ester with \textbf{1} (1,5-dibromo-2,6-dimethoxynaphthalene), using Pd$_2$(dba)$_3$ as catalyst, SPhos as ligand, and K$_2$CO$_3$ as base in a toluene/water/ethanol (5:1:1) mixture at \SI{383}{\kelvin}.
Demethylation of \textbf{2} with boron tribromide at \SI{273}{\kelvin}, afforded the dihydroxy ternaphthalene (\textbf{3'}) quantitatively (~99\%), which was subsequently converted into the pure bis(triflate)ternaphthyl derivative (\textbf{3}) in 72\% yield after purification by silica gel column chromatography (details on \textbf{3'} are provided in the Supporting Information). Following our previously published procedure \cite{sharma_peri-alkylated_2024}, bis(triflate)ternaphthyl derivative (\textbf{3}) was cross-coupled under Suzuki conditions with trimethyl(2'-(4,4,5,5-tetramethyl-1,3,2-dioxaborolan-2-yl)-[1,1'-biphenyl]-4-yl)silane (\textbf{1c}) to yield precursor \textbf{4} in 80\%. Compound \textbf{5} was then prepared by iodination of the TMS-protected \textbf{4} with iodine monochloride (ICl) at \qty{203}{\kelvin}, which cleanly converted the TMS groups into iodo functionalities, and afforded the desired \textbf{5} in \num{93}\% isolated yield. Maintaining the low reaction temperature was critical, as competing chlorination of the outer naphthalene unit was observed at temperatures above \qty{203}{\kelvin}.

The molecular structure of \textbf{5} was further confirmed by single-crystal X-ray diffraction. Crystals suitable for diffraction measurements were grown by slow vapor diffusion of hexane into a dichloromethane (DCM) solution. Structural analysis revealed that compound \textbf{5} crystallizes in the monoclinic P21/n space group. Additional details on the synthesis and characterization of the precursor molecules leading to \textbf{5} are provided in the Supporting Information.

\begin{figure}[h]
    \centering
    \includegraphics[width=1\linewidth]{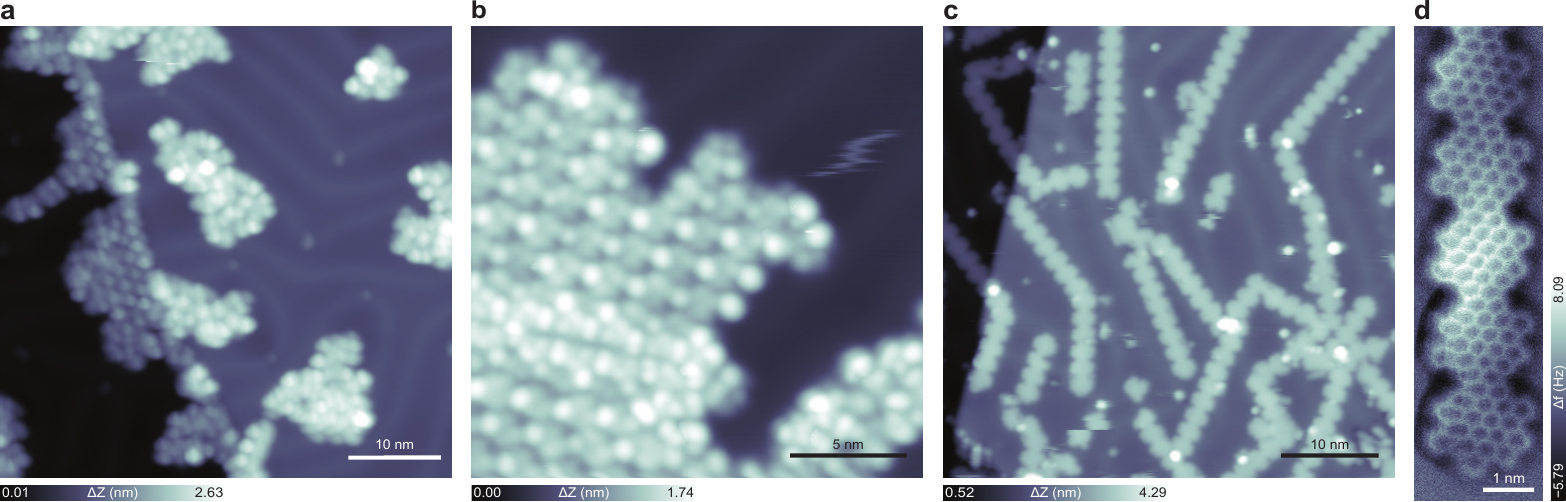}
    \caption{On-surface synthesis of \xGNR{} via surface-assisted dehalogenation and polymerization of monomer \textbf{5}, followed by cyclodehydrogenation of polymer \textbf{6}. 
    \textbf{(a–c)} STM topography image of monomers \textbf{5}, polymer \textbf{6}, and \xGNR{} \textbf{7} on the Au(111) surface. All STM measurements were performed at \SI{4.5}{\kelvin}. 
        Scanning parameters: (a) \SI{-0.4}{\volt}, \SI{20}{\pico\ampere}; 
        (b) \SI{1}{\volt}, \SI{50}{\pico\ampere}; 
        (c) \SI{-1}{\volt}, \SI{20}{\pico\ampere}. 
    The STM images were post‑processed by subtracting a plane fitted through three reference points.
    \textbf{(d)} High-resolution nc-AFM frequency shift image of \xGNR{} using a CO-functionalized tip with oscillation amplitude of 100 pm. The nc-AFM measurement was also performed at \SI{4.5}{\kelvin}.}
    \label{fig:STM_AFM}
\end{figure}

Compound \textbf{5} was subsequently sublimated at \SI{573}{\kelvin} onto an Au(111) surface held at room temperature under ultra-high vacuum (UHV). As shown in Figure~\ref{fig:STM_AFM}a, deposition of \textbf{5} produced islands of self-assembled intact molecules. Annealing the sample at \SI{438}{\kelvin} for \SI{30}{\minute} initiated surface-assisted dehalogenation followed by C–C coupling, yielding polymer \textbf{6} which itself organized into extended islands driven by attractive interactions between the partially interdigitated phenyl rings of neighboring chains (Figure~\ref{fig:STM_AFM}b)~\cite{talirz_-surface_2017,cai_atomically_2010}. A second annealing step at \SI{673}{\kelvin} for \SI{30}{\minute} thermally induced cyclodehydrogenation of the polymers, leading to the formation of the \xGNR{} \textbf{7}. The resulting ribbons are well defined and of high structural quality (Figure~\ref{fig:STM_AFM}c), and the bond-resolved nc-AFM image in  Figure~\ref{fig:STM_AFM}d confirms the atomically precise gulf-edged chiral structure characteristic of the \xGNR{}. A length statistics study of ribbons is provided in the Supporting Information.

\subsection{Electronic structure characterization of \xGNR{}}

\begin{figure}[h]
    \centering
    \includegraphics[width=1\linewidth]{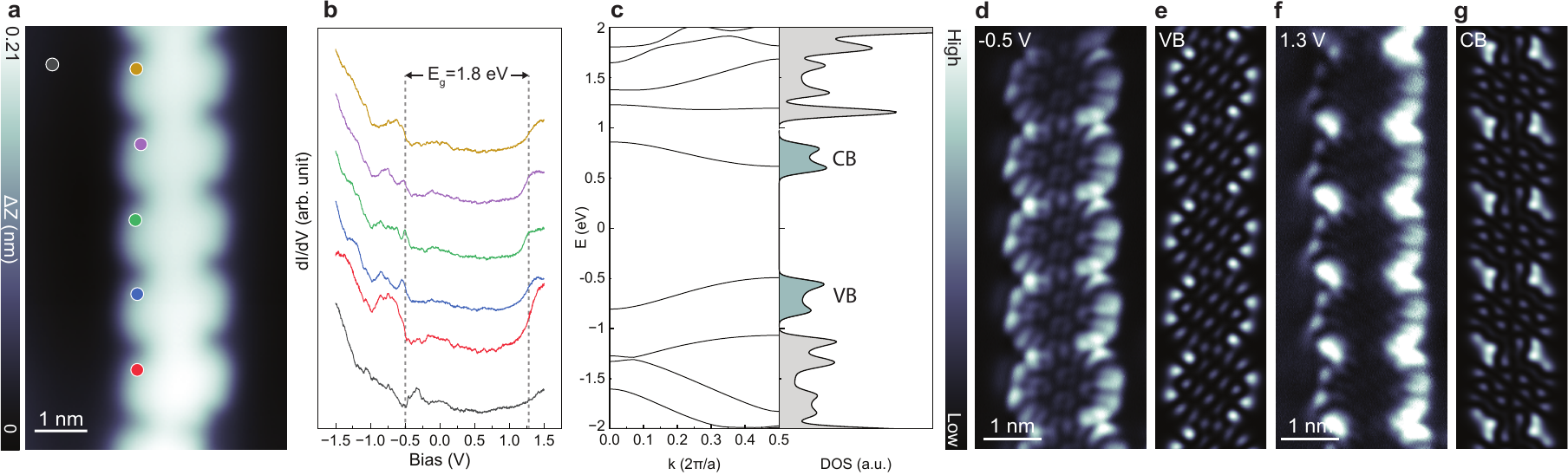}
    \caption{Electronic characterization of the \xGNR{}. 
    \textbf{(a,b)} The STS measurements in (b) were acquired on the 5-unit \xGNR{} segment shown in (a). The spectra are color-coded to their STS positions in (a). A CO-functionalized tip was used to acquire STS spectra. 
    \textbf{(c)} Band structure (energy versus wave vector k) (left, black; a is the lattice parameter) and density of states (right, grey) calculated for an infinitely long \xGNR{} with Kohn–Sham DFT. Valence band (VB) and conduction band (CB) are highlighted.
    \textbf{(d, f)} Constant current differential conductance (dI/dV) maps acquired over a 5-unit \xGNR{} segment at the energy of the valence band and conduction band. STS spectra and dI/dV maps were acquired on two different ribbons. A CO-functionalized tip was used to acquire the dI/dV maps. 
    \textbf{(e,g)} Simulated LDOS maps within the valence and conduction bands for an infinitely long \xGNR{}, calculated at a tip–sample distance of \SI{3.203}{\angstrom}.}
    \label{fig:dIdV_CB_VB}
\end{figure}

To determine the valence band (VB) and conduction band (CB) positions of the \xGNR{}, STS measurements were performed on a 5-unit \xGNR{} segment. The Shockley surface state on Au(111), visible near \SI{-0.5}{\volt}, served as an indicator of the tip quality~\cite{wang_automated_2021}. By comparing the differential conductance (dI/dV) spectra acquired on the \xGNR{} and the Au(111) substrate (Figure~\ref{fig:dIdV_CB_VB}b), we identify the VB maximum at \SI{-0.5}{\volt} and the CB minimum at \SI{1.3}{\volt}, corresponding to an experimental band gap of \SI{1.8}{\electronvolt}. The asymmetric alignment of these bands with respect to the Fermi level is attributed to the high work function of the Au(111) substrate, consistent with a previous study that investigated Fermi level pinning of GNRs on Au(111)~\cite{merino-diez_width-dependent_2017}. Comparison of the \SI{1.8}{\electronvolt} band gap of \xGNR{} with those reported for other chGNRs reveals a general trend in which the band gap decreases with increasing ribbon width, analogous to the behavior observed in armchair GNRs (given the different families)~\cite{keerthi_-surface_2020,liu_cove-edged_2024,merino-diez_unraveling_2018,deniz_electronic_2025,li_topological_2021}. However, the electronic properties of chGNRs are also strongly influenced by their edge topology and chirality, which can lead to significant deviations from this width-dependent trend.

To further validate this assignment, we acquired spatially resolved dI/dV maps of a 5-unit \xGNR{} segment at sample biases corresponding to the VB maximum and CB minimum (Figure~\ref{fig:dIdV_CB_VB}d,f). These experimental maps agree well with simulated local density of states (LDOS) distributions obtained from periodic DFT (see Methods), supporting the assignment of the VB and CB energies. The simulation converges to a closed-shell configuration with a band gap of \SI{1.1}{\electronvolt}, indicating that the \xGNR{} does not host any spin-localized state, despite the presence of zigzag edges (Figure~\ref{fig:dIdV_CB_VB}c).

The discrepancy between the experimental band gap of \SI{1.8}{\electronvolt} and the DFT-calculated band gap of \SI{1.1}{\electronvolt} can be attributed to three factors. First, the measurements were performed on an Au(111) substrate, whereas the calculations were carried out for a freestanding GNR. The screening effect of the metallic substrate can reduce the band gap of the GNR~\cite{deniz_revealing_2017,ruffieux_electronic_2012,neaton_renormalization_2006}. Second, defects at the ends of the 5-unit (4,2,7)-chGNR segments used for STS and dI/dV measurements can induce quantum-confinement effects, resulting in a larger experimental band gap than the intrinsic value. Third, while DFT reliably captures orbital shapes and energy ordering, it is known to systematically underestimate band gaps in GNRs~\cite{louie_chapter_2006}. Because these three factors contribute to the discrepancy between the experimental and DFT-calculated band gaps in different ways, the observed discrepancy cannot be quantitatively attributed to any single factor.

\subsection{Raman characterization of \xGNR{}}

\begin{figure}[h]
    \centering
    \includegraphics[width=1\linewidth]{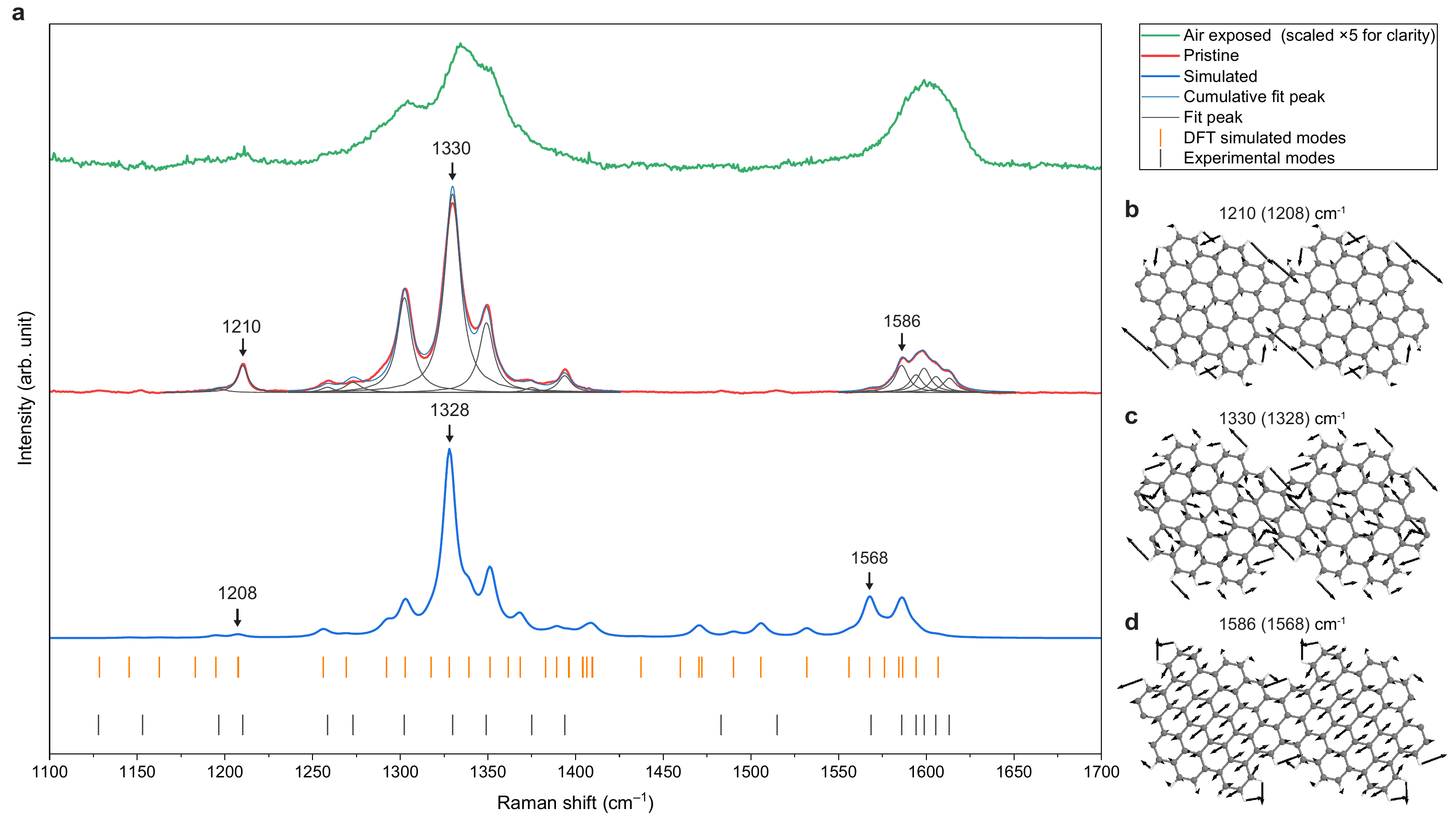}
    \caption{High-frequency region of the Raman spectrum for \xGNR{} and the DFT-simulated vibrational modes. (a) Experimental Raman spectra of pristine and air-exposed \xGNR{} under \SI{532}{\nano\meter} excitation, along with the DFT-simulated Raman spectrum. The simulated spectrum is obtained by summing Lorentzian peaks with a broadening of \SI{9}{\per\centi\meter} at the calculated vibrational frequencies. In the experimental spectrum of pristine \xGNR{}, the D and G peak regions are deconvoluted using Lorentzian fits. Vertical lines beneath the spectra show a comparison between experimentally determined Raman peak positions obtained through deconvolution, and those obtained from DFT simulation. (b, c, d) Visualization of three vibrational modes of \xGNR{} based on DFT simulation. The vibrational frequency and the corresponding simulated frequency (in parentheses) are shown at the top of each panel.}
    \label{fig:Raman_627a}
\end{figure}

The Raman spectrum of GNRs typically consists of two main regions: a low-frequency region, including the longitudinal compressive mode (LCM) and the radial breathing-like mode (RBLM), and a high-frequency region, characterized by the  CH/D modes and G mode. In the low-frequency region, the LCM appears around \SI{100}{\per\centi\metre} and is sensitive to the GNR length~\cite{overbeck_universal_2019}, while the RBLM serves as a unique fingerprint of the GNR width. For example, the RBLMs for 5-, 7-, 9-, and 17-AGNRs are reported at \SI{529}{\per\centi\metre}, \SI{398}{\per\centi\metre}, \SI{311}{\per\centi\metre}, and \SI{169}{\per\centi\metre}, respectively~\cite{overbeck_optimized_2019,hwang_optimized_2025}. In the high-frequency region, CH/D modes located in the \SIrange{1100}{1500}{\per\centi\metre} range are associated with intrinsic backbone vibrational modes (zone-folded LO/TO/LA phonons) along with contributions from C-H vibrations~\cite{darawish_role_2025,verzhbitskiy_raman_2016, nascimento_optical_2025}. The most prominent Raman-active feature, the G mode, appears around \SI{1600}{\per\centi\metre}; it is present in all sp\textsuperscript{2} carbon-based materials and originates from in-plane bond stretching vibrations~\cite{darawish_role_2025,verzhbitskiy_raman_2016}.

To characterize the vibrational properties of pristine \xGNR{}, Raman spectroscopy was performed under UHV on a high‑coverage \xGNR{} sample deposited on a \SI{4}{\milli\meter} $\times$ \SI{4}{\milli\meter} Au/mica substrate using a \SI{532}{\nano\meter} excitation wavelength. A large‑scale STM image of this sample is provided in the Supporting Information. The experimental spectrum represents an average over a \SI{50}{\micro\meter} $\times$ \SI{50}{\micro\meter} area, sampled at 50 $\times$ 50 points. Peaks in the low-frequency region are too weak to be resolved, whereas the high-frequency region shows pronounced features in both the CH/D and G mode ranges (Figure~\ref{fig:Raman_627a}a, red spectrum;  a full Raman spectrum is provided in the Supporting Information). Compared with AGNRs, \xGNR{} exhibits substantially more peaks in these regions, reflecting the additional symmetry breaking of its complex structure. Peak deconvolution is therefore required to assign the vibrational modes. 

To guide the deconvolution, we performed a non-resonant Raman simulation based on periodic DFT under conditions approximating the experimental, assuming randomly oriented \xGNR{}s on a two-dimensional plane (see Methods). All modes in the high-frequency range exhibit $A_g$ symmetry. The simulation does not account for substrate effects or resonance contributions, which may cause deviations in peak intensities and frequencies~\cite{overbeck_optimized_2019}. The simulated spectrum was generated by summing Lorentzian functions of \SI{9}{\per\centi\meter} broadening centered at the calculated vibrational frequencies (Figure~\ref{fig:Raman_627a}a, blue spectrum). The full simulated Raman activity spectrum is provided in the Supporting Information.

Guided by the simulation, we fitted the experimental high-frequency region with Lorentzian functions, with the full width at half maximum (FWHM) constrained between 7–\SI{11}{\per\centi\meter}, and assigned each peak to a specific vibrational mode.  In the CH/D region, the peak at \SI{1210}{\per\centi\meter} corresponds to a simulated mode at \SI{1208}{\per\centi\meter}, primarily involving C–H vibrations at the gulf edge and therefore serves as a structural fingerprint for this edge topology. The most intense peak in this region, at \SI{1330}{\per\centi\meter} (simulated at \SI{1328}{\per\centi\meter}), reflects a mixed vibration of edge C–H bonds and the C-C bonds in the GNR backbone. In the G region, the frequency underestimation by DFT is more pronounced than in lower frequencies, a known signature of DFT's inability to capture many-body effects such as strong electron-phonon coupling~\cite{piscanec_kohn_2004}. For example, the peak at \SI{1586}{\per\centi\meter} corresponds to a simulated mode at \SI{1568}{\per\centi\meter}, dominated by C-C bond stretching in the GNR backbone (Figure~\ref{fig:Raman_627a}d). Comparing the three vibrational modes illustrated in Figure~\ref{fig:Raman_627a}b–d, lower-frequency modes are dominated by edge C–H vibrations, whereas higher-frequency modes involve the carbon backbone.

Beyond identifying the vibrational modes, Raman spectroscopy also provides a sensitive probe of ambient stability, a critical requirement for the practical implementation of GNR-based devices. Zigzag-edge segments of GNRs have been shown to exhibit high reactivity when exposed to oxygen or air, even when the structures have a predominantly closed-shell character~\cite{berdonces-layunta_chemical_2021}. For instance, teranthene and hexanthene, both with a high zigzag-to-armchair edge ratio, show pronounced broadening of the CH, D, and G modes after only five minutes of exposure~\cite{barin_-surface_2023}. (3,1,4)-chGNR similarly degrades under oxygen exposure, with STM and AFM revealing that oxidation is regioselective to its zigzag segments~\cite{berdonces-layunta_chemical_2021}, and the zigzag termini of 5AGNR also undergo structural changes under oxygen exposure~\cite{lawrence_probing_2020}. In contrast, AGNRs with low zigzag-to-armchair edge ratios or purely armchair edges, such as 7- and 9-AGNR, exhibited excellent long-term stability in air, with no significant broadening of their Raman peaks even after several months of exposure~\cite{borin_barin_surface-synthesized_2019,fairbrother_high_2017}.

The Raman response of \xGNR{} upon air exposure is consistent with this picture. Figure~\ref{fig:Raman_627a}a (green spectrum) shows the Raman spectrum of \xGNR{} on an Au substrate after \SI{30}{\minute} of air exposure. This short exposure severely degraded the D and G peaks, rendering peak deconvolution and FWHM analysis highly challenging due to the absence of well-resolved features. A direct comparison between the pristine and air-exposed spectra reveals that the CH and D peaks are far more suppressed than the G peak, suggesting that the structural damage occurs both within the lattice and at the edges. These observations are consistent with previous reports that ambient instability can persist in GNRs containing zigzag edge segments, even when the overall electronic structure is predominantly closed-shell~\cite{berdonces-layunta_chemical_2021}. Our findings further highlight the need for additional studies on the ambient stability of chGNRs, even when no open-shell characteristics are observed.

\section{Conclusion}
We have demonstrated a new precursor design strategy for the on-surface synthesis of a gulf-edged chiral GNR, the \xGNR{}. By introducing biphenyl substituents at the central naphthalene unit of a trisnaphthalene core, the monomer imposes conformational constraints that suppress undesired side reactions and enhance surface anchoring through its extended $\pi$-conjugation, thereby enabling clean polymerization and cyclodehydrogenation up to \SI{673}{\kelvin}. STM and nc-AFM confirm that the resulting ribbons are long, regiospecific, and atomically precise, with the expected gulf-edged chiral structure. STS  measurements combined with DFT simulations identify the \xGNR{} as a closed-shell semiconductor with an experimental band gap of \SI{1.8}{\electronvolt}. Raman spectroscopy further characterizes the vibrational properties of the \xGNR{}, revealing 
a distinctive C-H vibration at the gulf edge that can serve as a structural fingerprint for this edge topology. Air-exposure measurements show, however, that the \xGNR{} is unstable under ambient conditions despite its large band gap and closed-shell electronic structure, consistent with previous reports that zigzag edge segments can drive ambient degradation independently of the overall spin configuration.

\section{Methods}
\label{sec_me}

\subsection{Monomer synthesis}
Detailed descriptions of the reaction steps and characterization of precursor molecules including single-crystal analysis (CCDC 2537291) of compound 5 are provided in the Supporting Information.

\subsection{On-surface synthesis}
On-surface synthesis steps were performed in a UHV chamber, directly connected to a low-temperature scanning tunneling microscope (Scienta Omicron). Au(111) thin films epitaxially grown on mica (Phasis Sàrl, Geneva, Switzerland) were used as growth substrates in this work. Au(111) on mica was cleaned by two cycles of argon ion sputtering at a pressure of \SI{8e-6}{\milli\bar} for \SI{10}{\minute}, followed by annealing to \SI{723}{\kelvin} for \SI{10}{\minute}. Before surface deposition, the precursor \textbf{5} in powder form was thoroughly degassed at its sublimation temperature of \SI{573}{\kelvin}, with the deposition rate monitored with a quartz microbalance. During deposition, the Au(111) substrates were kept at room temperature, around \SI{298}{\kelvin}. After deposition, the samples were heated to \SI{438}{\kelvin} for \SI{30}{\minute} and \SI{673}{\kelvin} for \SI{30}{\minute} to induce polymerization and cyclodehydrogenation. The temperature of the Au(111) on mica was measured with an infrared pyrometer (Optris, Berlin, Germany). 

\subsection{Scanning probe microscopy measurements}
STM, STS, and nc-AFM were performed in commercial low-temperature STM/nc-AFM systems from Scienta Omicron, operating at a temperature of \SI{4.5}{\kelvin} and a base pressure below \SI{2e-11}{\milli\bar}. All STS spectra, dI/dV maps, and the nc-AFM image were acquired with a CO-functionalized tungsten tip. In situ cold deposition of CO molecules was performed to obtain a CO-functionalized tip. STS spectra and dI/dV maps were acquired using a lock‑in technique with a modulation frequency of \SI{681}{\hertz}. A modulation amplitude of \SI{10}{\milli\volt} was used for all the STS spectra and \SI{20}{\milli\volt} for all the dI/dV maps. The nc-AFM image was taken with a qPlus tuning fork sensor (resonance frequency, \(\sim\)\SI{25.4}{\kilo\hertz}; quality factor, \(\sim\)14\,000)~\cite{giessibl_qplus_2019}. A CO-functionalized tungsten tip was used, and measurements were conducted in the constant-height mode.

\subsection{Raman spectroscopy}
To measure the Raman spectra of pristine \xGNR{}s, a high‑coverage sample was first prepared and characterized by STM, then transferred into a home-built vacuum chamber equipped with optical access for the Raman microscope. The chamber was pumped by an ion pump and maintained at a pressure below \SI{1e-6}{\milli\bar}, and was connected directly to the fast‑entry lock of the STM system, allowing the sample to remain under UHV conditions throughout the transfer and throughout the Raman measurement. Following the UHV measurement, the sample was removed, exposed to ambient air for \SI{30}{\minute}, and then transferred to a different vacuum chamber at a pressure of approximately \SI{1e-5}{\milli\bar} for Raman characterization of the air-exposed \xGNR{}s.

Raman measurements were carried out in a backscattering geometry using a Witec Alpha~300~R confocal Raman microscope equipped with a 532~nm laser line operating at a power of \SI{30}{\milli\watt} and an 1800~g/mm grating. A 50$\times$ LD objective (Zeiss, NA = 0.55) was used to focus the laser beam onto the sample and to collect the scattered light. For each spectrum, a Raman mapping approach was employed: 2500 individual spectra (1~s integration time) were recorded over a grid of 50~$\times$~50 points (\SI{50}{\micro\meter}~$\times$~\SI{50}{\micro\meter}), and the average of these spectra was used for analysis. The Raman data were processed in OriginPro.

\subsection{Computational methods}
The band structure and Raman spectrum calculations were performed with the DFT code Q\textsubscript{UANTUM} ESPRESSO 7.4~\cite{Giannozzi_2009,Giannozzi_2017,Giannozzi2020} via AiiDAlab~\cite{aiidalab}, using the AiiDAlab-QE application~\cite{AiiDAlabQE} together with the vibroscopy plugin AiiDAlab-qe-vibroscopy. This plugin integrates automated AiiDA-based~\cite{aiida} workflows for phonon and Raman calculations based on finite-displacement and finite-field approaches (aiida-vibroscopy)~\cite{bastonero_automated_2024}. The calculations employed the Perdew–Burke–Ernzerhof (PBE) exchange–correlation functional~\cite{perdew_Generalized_1996}. Plane-wave cutoffs of 100 Ry and 400 Ry were used for the wavefunctions and charge density, respectively. The ionic cores were described using pseudopotentials from the SSSP PBE precision library (version 1.3)~\cite{sssp_1.3}. The Brillouin zone was sampled using a 5 $\times$ 1 $\times$ 1 Monkhorst–Pack grid.

\section*{Supporting information}

\begin{itemize}
   \item Nomenclature of chGNRs; detailed procedures for the organic synthesis, including NMR spectra of all compounds and intermediates; electronic characterization of the VB‑1 and CB+1 of \xGNR{}; constant‑height dI/dV maps of \xGNR{}; a large‑scale STM topography image of the sample used for the Raman measurements; a statistical study of GNR lengths based on the large-scale STM topography image; complete simulated and experimental Raman spectra of \xGNR{}; and visualizations of the vibrational modes of \xGNR{} in the D and G bands.
\end{itemize}

\section*{Acknowledgements}

X.L., A.K., R.S., R.F., and G.B.B. greatly appreciate the financial support from the Werner Siemens Foundation (CarboQuant). X.L., R.S., M.L.P., R.F., and G.B.B. acknowledge funding by the European Union’s Horizon Europe research and innovation program under grant agreement no. 101099098 (ATYPIQUAL) and the SERI under contract number 23.00422. G.B.B. acknowledges funding from the Swiss National Science Foundation grant no. 200021E-219172/1 (GRAAL). This work is supported by a Royal Society International Exchanges grant IES\textbackslash R3\textbackslash 243222. The authors would like to thank Diamond Light Source for beamtime (proposal cy31541), and the staff of beamline I19 for their assistance. A.O-G, and C.P. acknowledge support from the NCCR MARVEL, a National Centre of Competence in Research, funded by the Swiss National Science Foundation (Grant number 205602). DFT calculations were performed at the Swiss National Supercomputing Centre (CSCS) under project ID lp83. M.L.P. acknowledges funding from SNSF for the Eccellenza Professorial Fellowship no. 203663 and project 10004856, as well as support by the Swiss State Secretariat for Education, Research and Innovation (SERI) under contract number MB22.00076 (ERC Starting Grant E-CONVERT). This work is partially supported by EPSRC Materials for Quantum Network (EP/W037912/1). The authors thank Dr. Chenxiao Zhao, Dr. Mirjana Dimitrievska, Dr. Wenze Gao, and Dr. Nicolo Bassi for helpful discussions.

\printbibliography

\clearpage

\begin{figure}[h]
    \centering
    \includegraphics[width=0.5\linewidth]{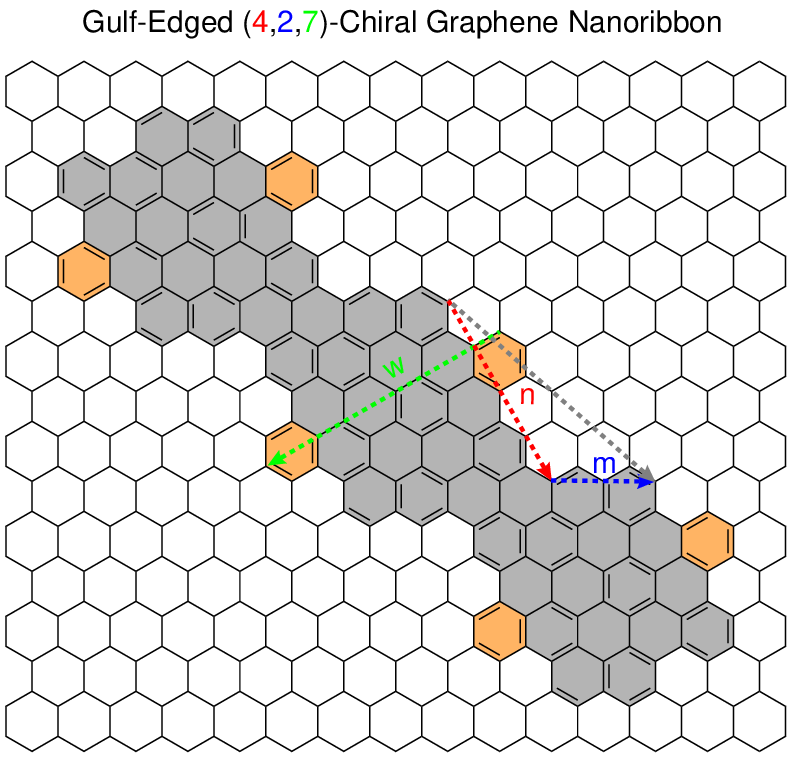}
    \caption{For Table of Contents Only}
    \label{fig:placeholder}
\end{figure}

\clearpage



\section*{Supplementary material}

# Contents

# 1 Nomenclature of chiral graphene nanoribbons

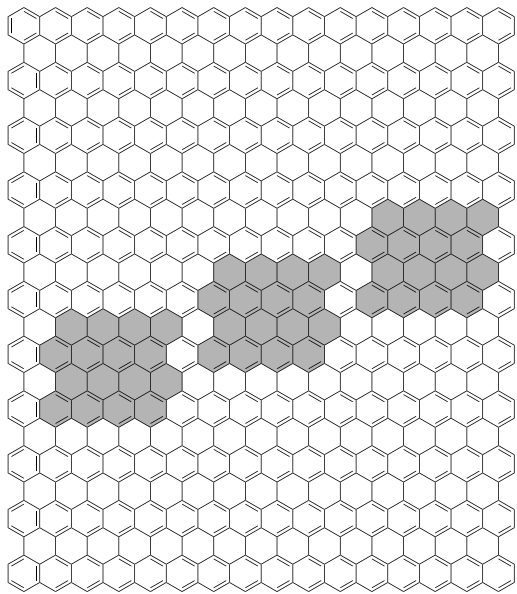

Nanographene building blocks used to construct the ribbon.

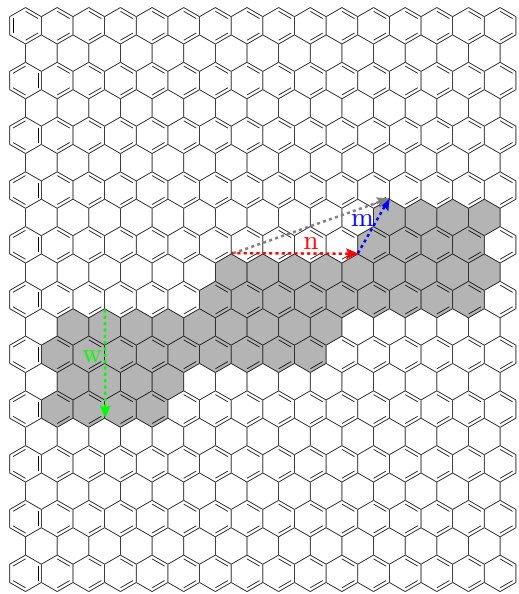

Structure of the resulting (4,2,5) chGNR, showing the chiral vector $(n,m)$ and the width w measured perpendicular to the zigzag edges.

Figure S1: Nomenclature of a (4,2,5) chiral graphene nanoribbon.

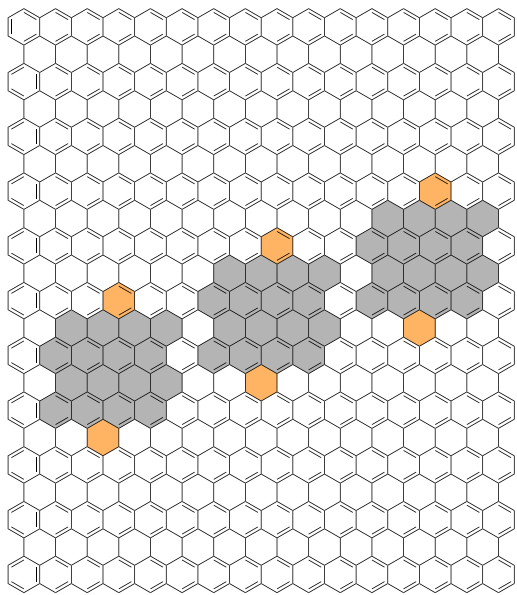

Nanographene building blocks used to construct the gulf-edged ribbon, with the additional benzenoid rings highlighted.

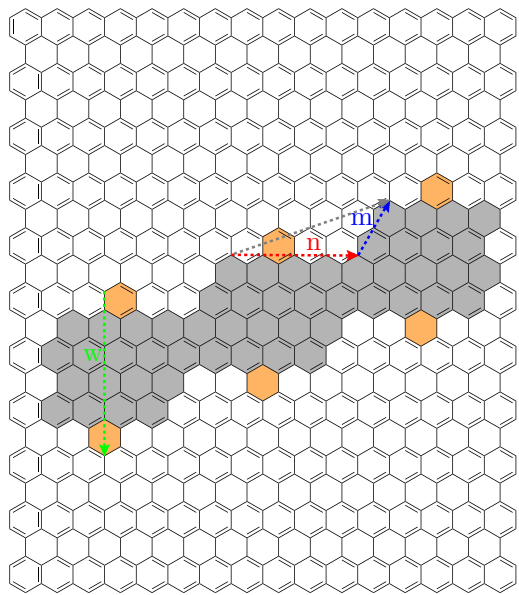

Structure of the resulting gulf-edged (4,2,7) chGNR, showing the chiral vector $(n,m)$ and the width w measured perpendicular to the zigzag edges at the position where the maximum width is defined.

Figure S2: Nomenclature of the gulf-edged (4,2,7) chiral graphene nanoribbon.

Chiral graphene nanoribbons (chGNRs) are generally defined by a chiral vector $\mathbf{C}_\mathrm{h}(n,m)$ ($n > m$), where $n$ and $m$ are indices of the graphene unit vectors [1–7]. However, the chiral vector alone does not fully characterize the structure of chGNRs. For example, chGNRs with different widths can share the same chiral vector.

As more distinct chGNRs have been synthesized, it has become necessary to refine the nomenclature to more clearly distinguish different structures. The ribbon width is therefore now included in the chGNR designation. To date, most chGNRs have been constructed by fusing nanographene building blocks along their armchair edges, with defined lateral shifts between adjacent blocks. In such cases, the width $w$ is defined perpendicular to the zigzag edges [8, 9]. Figure S1 illustrates how the name of a (4,2,5) chGNR is defined, following the $(n, m, w)$ notation where $w$ denotes the ribbon width. In this work, we follow the same convention.

However, the building block of the gulf-edged (4,2,7) chGNR contains an additional benzenoid ring on each side, so that the measurable width varies along the ribbon. To keep the nomenclature unambiguous, we define it at the position where the width reaches its maximum. Figure S2 shows how we assign the name of the gulf-edged (4,2,7) chGNR in this work.

Note that the "building block" referred here denotes only the geometric unit used to define the nomenclature and is not equivalent to the molecular precursor used in the on-surface synthesis of chGNRs.

# 2 Synthetic procedures towards compound 5 and characterizations

## 2.1 General methods

Unless otherwise specified, the reactions were carried out under an argon atmosphere in flame-dried glassware. All commercially available chemicals were purchased from Merck, Sigma-Aldrich, Fluorochem, or Alfa Aesar and used as received unless specified otherwise. Analytical thin layer chromatography (TLC) was performed on silica gel plates (Merck 60F254) visualized with a UV lamp (254 nm). Flash column chromatography was performed over silica gel (60–230 mesh). Organic extracts were dried over anhydrous $Na_2SO_4$ or $MgSO_4$. All NMR spectra were recorded on a Bruker AVANCE 500 or 400 at room temperature. $^1$H NMR spectra were recorded in $CDCl_3$ and the observed signals are reported as follows: chemical shift in parts per million from tetramethylsilane with the solvent as an internal indicator ($CDCl_3$ $\delta$ 7.26 ppm), multiplicity (s = singlet, d = doublet, t = triplet, q = quartet, p = pentet, m = multiplet or overlap of non-equivalent resonances, br = broad), integration. $^{13}$C NMR spectra were recorded at 126 or 101 MHz in $CDCl_3$, and the observed signals were reported as follows: chemical shift in parts per million from tetramethylsilane with the solvent as an internal indicator ($CDCl_3$ $\delta$ 77.16 ppm). Atmospheric Pressure Chemical Ionization (APCI) high-resolution mass spectrometry was performed using a Thermo Orbitrap Exactive Plus Extended Mass Range mass spectrometer to detect the molecular ion plus hydrogen ($[M+H]^+$).

## 2.2 Synthesis of compound 1c

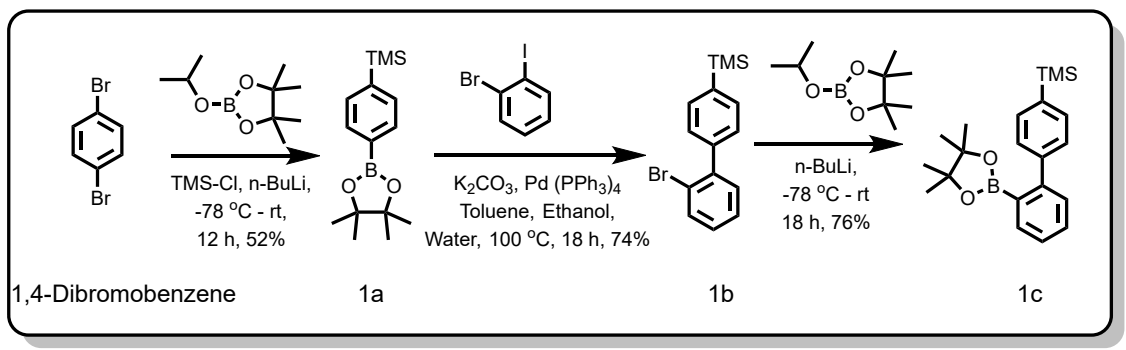

Figure S3: Synthetic scheme for **1c**.

Figure S3 illustrates the synthesis of **1c**, one of the precursors for the synthesis of **4**. Two bromo groups of 1,4-dibromobenzene were selectively replaced with trimethylsilyl (TMS), followed by boronic ester groups. Initially, 1,4-dibromobenzene was dissolved in THF, and the solution temperature was reduced to $-78\,°C$, after which n-BuLi was added dropwise. The reaction mixture was stirred at this temperature for 30 min, after which TMSCl was added. The reaction mixture was stirred at room temperature for the subsequent 30 min. The reaction mixture was again cooled to $-78\,°C$, followed by the addition of n-BuLi and 2-isopropoxy-4,4,6-trimethyl-1,3,2-dioxaborinane, and stirred for 12 hours at room temperature, yielding **1a**. The TMS-protected phenylboronic ester **1a** was selectively coupled with 1-bromo-2-iodobenzene under Suzuki cross-coupling reaction conditions to yield biphenyl **1b**. The bromo group of biphenyl **1b** was transformed into the corresponding boronic ester utilizing n-BuLi and 2-isopropoxy-4,4,6-trimethyl-1,3,2-dioxaborinane reaction conditions, resulting in the synthesis of **1c** with a good yield.

5

### 2.2.1 Synthesis and characterization of intermediate 1a

trimethyl(4-(4,4,5,5-tetramethyl-1,3,2-dioxaborolan-2-yl)phenyl)silane (**1a**)

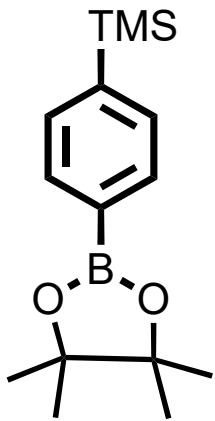

1,4-Dibromobenzene (4 g, 16.95 mmol) was solubilized in anhydrous THF (40 mL) under a nitrogen atmosphere and subsequently cooled to −78 °C. n-BuLi (11.1 mL, 1.6 M in hexane, 17.8 mmol) was added dropwise, and the reaction mixture was stirred for 10 min. Subsequently, Me$_3$SiCl (2.3 mL, 17.8 mmol) has been added dropwise, and the reaction was allowed to warm to room temperature before being cooled once more to −78 °C. Subsequently, an additional portion of n-BuLi (11.6 mL, 1.6 M in hexane, 18.64 mmol) was introduced dropwise. Following a 20 min stirring period at this temperature, dioxaborolane (2.76 mL, 13.56 mmol) was added dropwise. The reaction was allowed to reach room temperature, stirred overnight (12 hours), and subsequently quenched with the addition of water. The reaction mixture was poured into water and extracted with DCM, dried over MgSO$_4$, and the solvents were removed under reduced pressure. The crude product was purified by column chromatography with hexane:DCM (2:1) as eluent, yielding a white solid **1a** (2.43 g, 8.81 mmol, yield 52%).

**APCI-HRMS**: $m/z$ 276.1717, calculated 276.1790. **$^1$H NMR** (500 MHz, CDCl$_3$) $\delta$ 7.79 (d, $J$ = 7.8 Hz, 2H), 7.54 (d, $J$ = 7.9 Hz, 2H), 1.35 (s, 12H), 0.27 (s, 9H). **$^{13}$C NMR** (126 MHz, CDCl$_3$) $\delta$ 144.38, 134.00, 132.74, 83.88, 24.98, -1.10.

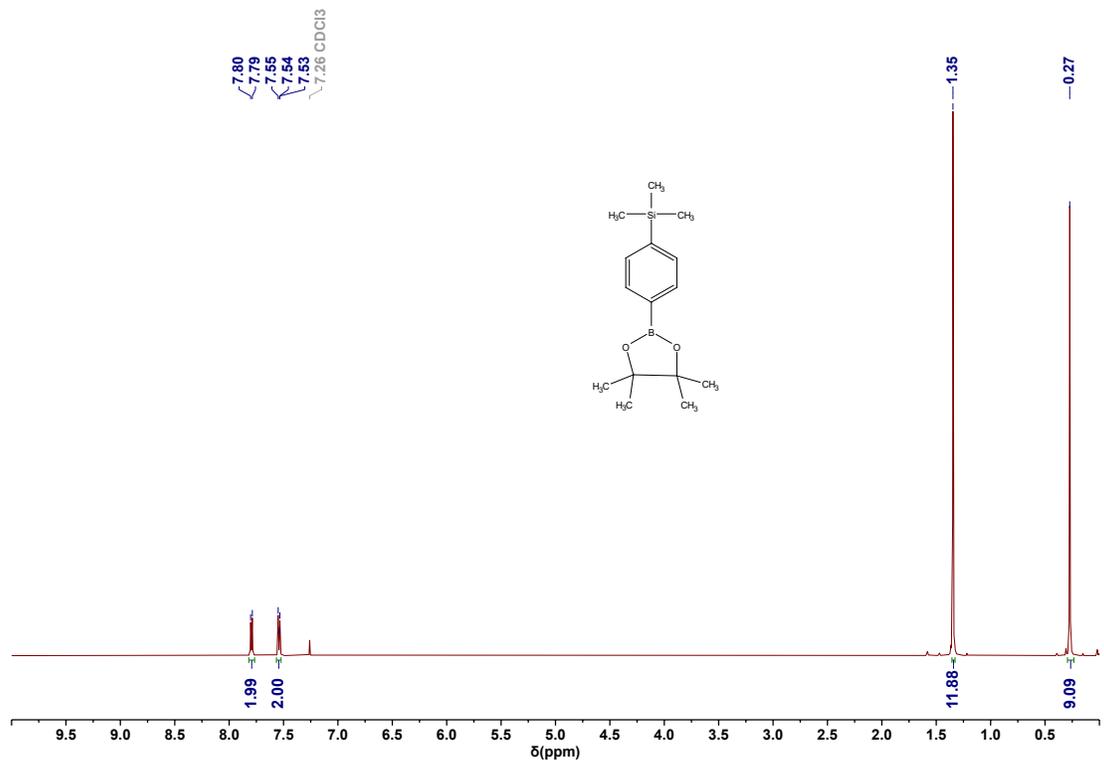

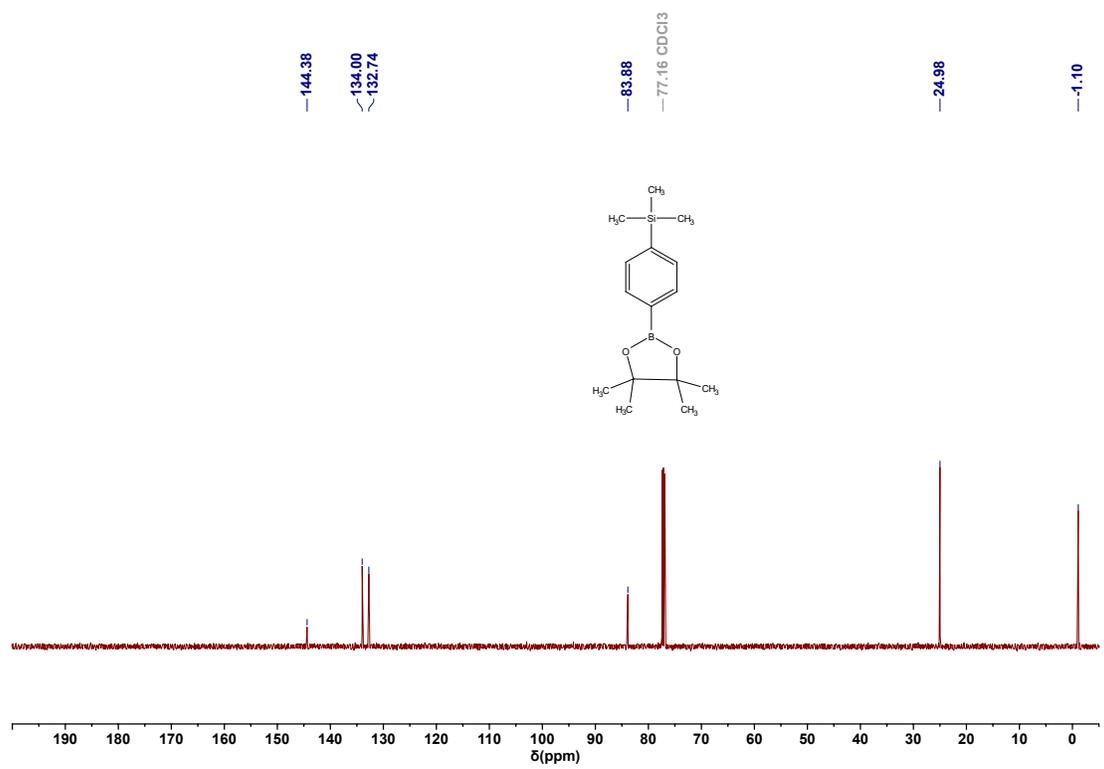

Figure S4: **¹H NMR** and **¹³C NMR** spectra of compound **1a**.

7

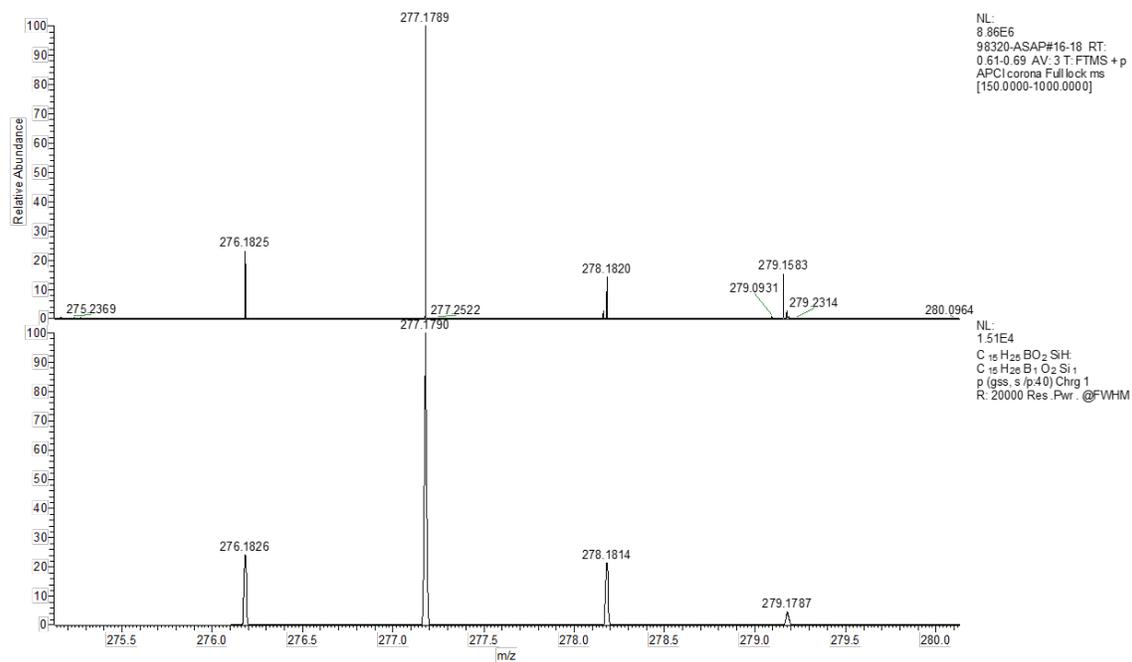

Figure S5: **APCI-HRMS** analysis for compound **1a**.

### 2.2.2 Synthesis and characterization of intermediate 1b

(2'-bromo-[1,1'-biphenyl]-4-yl)trimethylsilane (**1b**)

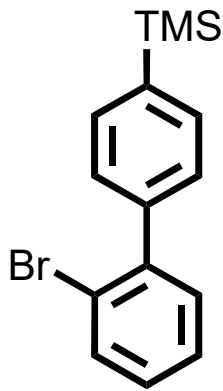

Compound **1a** (1.0 eq, 2.0 g, 13.3 mmol), 1-bromo-2-iodobenzene (1.3 eq, 3.7 g, 13.39 mmol), and $K_2CO_3$ (3.0 eq, 4.3 g, 30.9 mmol) were added to 40 mL of degassed toluene and 10 mL of degassed ethanol/$H_2O$ (5:5) under a nitrogen atmosphere. Following a 20 min degassing with nitrogen, the catalyst $Pd(PPh_3)_4$ (0.04 eq, 476 mg, 0.4 mmol) was added. After an additional 20 min of nitrogen degassing, the mixture was heated at 100 °C for 18 h. Following filtration of the reaction mixture through celite, the solvent was evaporated. The crude product was then purified via flash column chromatography employing a hexane: DCM (10: 1) eluent, resulting in a white solid (3.02 g, 9.884 mmol, yield 74%).

**APCI-HRMS**: $m/z$ 304.0252, calculated 304.0283. $^1$**H NMR** (500 MHz, CDCl$_3$) δ 7.68 (dd, $J$ = 7.9, 1.2 Hz, 1H), 7.59 (d, $J$ = 8.1 Hz, 2H), 7.41 (d, $J$ = 8.1 Hz, 2H), 7.38–7.31 (m, 2H), 7.20 (m, 1H), 0.32 (s, 9H). $^{13}$**C NMR** (126 MHz, CDCl$_3$) δ 142.67, 141.53, 139.87, 133.29, 133.12, 131.48, 128.83, 128.77, 127.53, 122.69, -0.92.

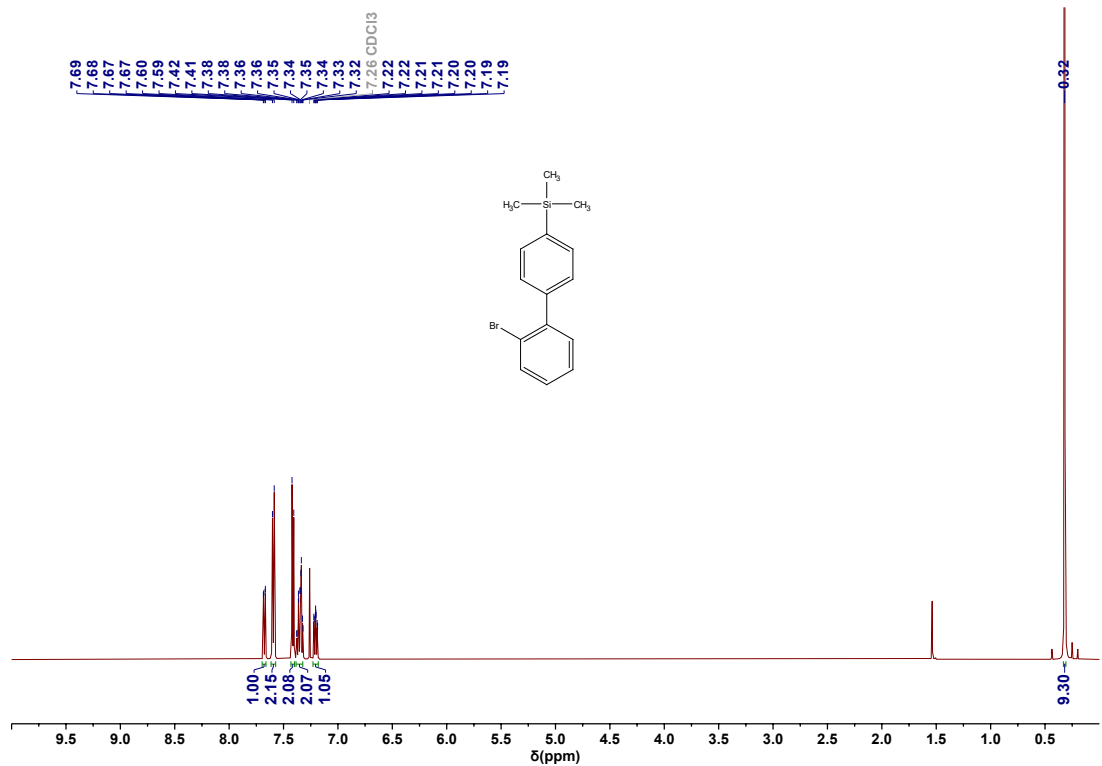

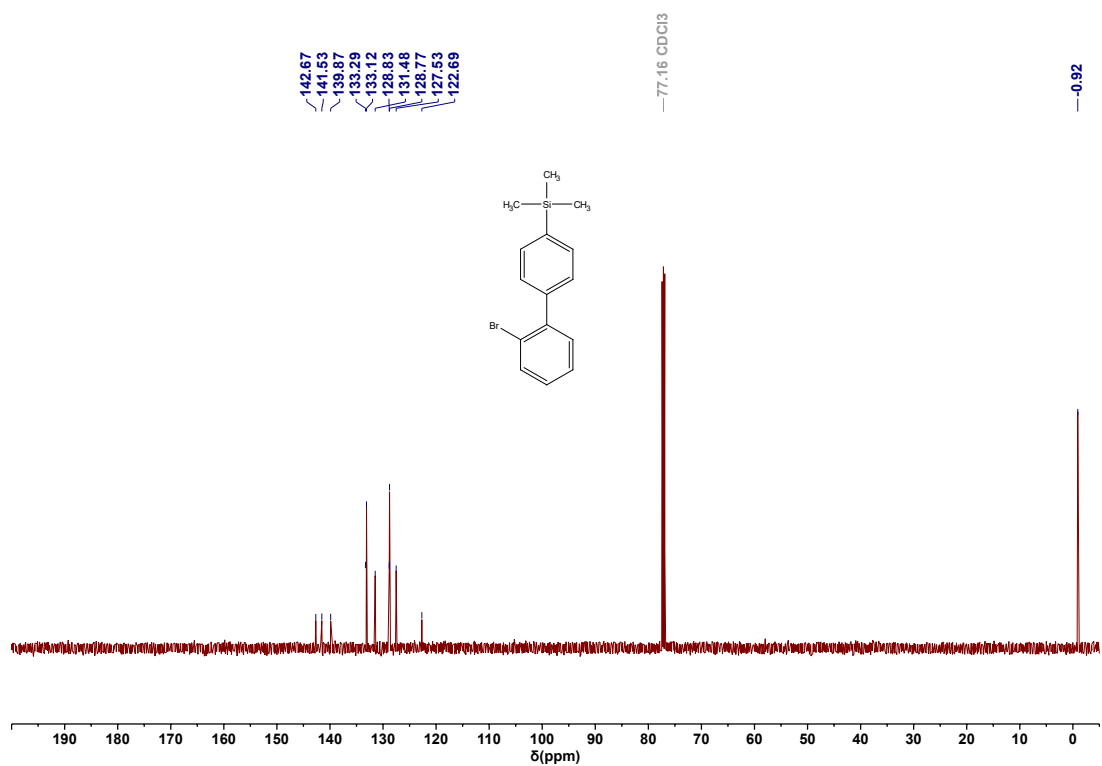

Figure S6: **¹H NMR** and **¹³C NMR** spectra of compound **1b**.

10

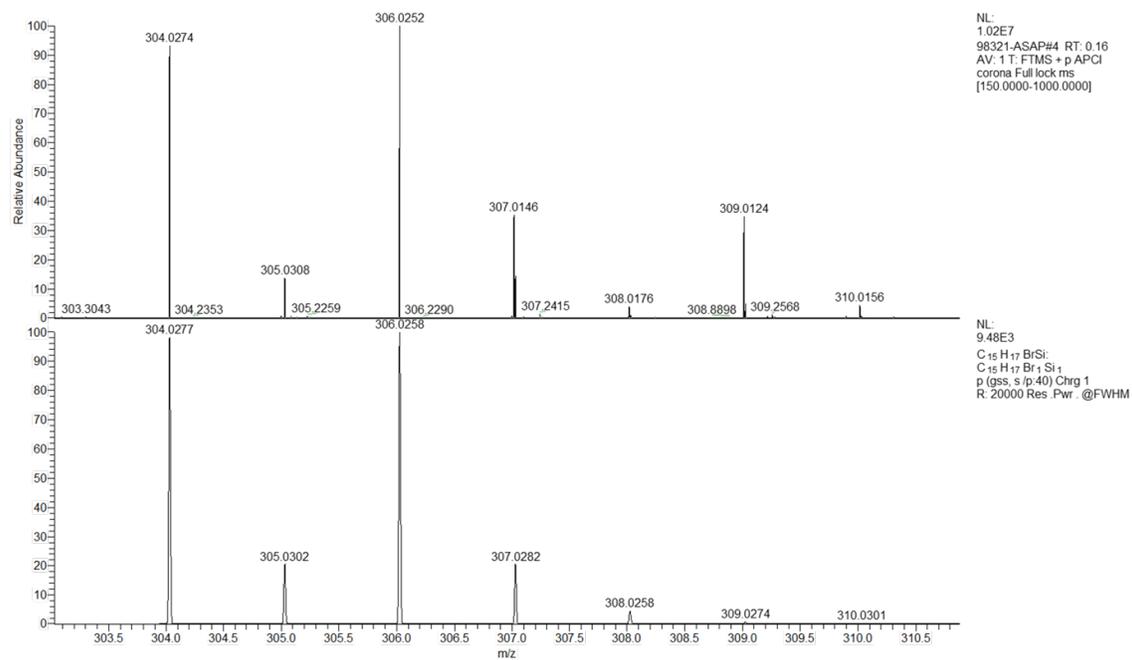

Figure S7: APCI-HRMS analysis for compound 1b.

### 2.2.3 Synthesis and characterization of intermediate 1c

Trimethyl(2'-(4,4,5,5-tetramethyl-1,3,2-dioxaborolan-2-yl)-[1,1'-biphenyl]-4-yl)silane (**1c**)

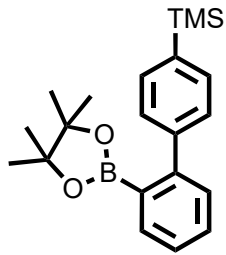

Compound **1b** (600 mg, 1.9 mmol) was solubilized in 20 mL of anhydrous THF under a nitrogen atmosphere. The solution was cooled to $-78\,^\circ$C prior to the dropwise addition of 1.8 mL (2.9 mmol) of n-BuLi (1.6 M solution in hexane). The reaction mixture was stirred for 20 min prior to the addition of 0.6 mL (2.9 mmol) of 2-isopropoxy-4,4,5,5-tetramethyl-1,3,2-dioxaborolane. The reaction mixture was allowed to equilibrate to ambient temperature and stirred for a further 12 h prior to being poured into ice water. The solution was extracted with diethyl ether; the organic phase was washed with brine and dried over $MgSO_4$ prior to solvent removal. The crude product was then purified via flash column chromatography employing a hexane: DCM (5: 1) eluent, resulting in a colorless oil (510 mg, 1.45 mmol, yield 76%).

**APCI-HRMS**: $m/z$ 352.2027, calculated 352.2030. **$^1$H NMR** (500 MHz, CDCl$_3$) $\delta$ 7.73–7.70 (m, 1H), 7.56–7.52 (m, 2H), 7.48–7.43 (m, 1H), 7.41–7.37 (m, 3H), 7.36–7.32 (m, 1H), 1.21 (s, 12H), 0.31 (s, 9H). **$^{13}$C NMR** (126 MHz, CDCl$_3$) $\delta$ 147.60, 143.79, 138.66, 134.47, 132.95, 130.17, 128.91, 128.58, 126.43, 83.85, 24.72, -0.91.

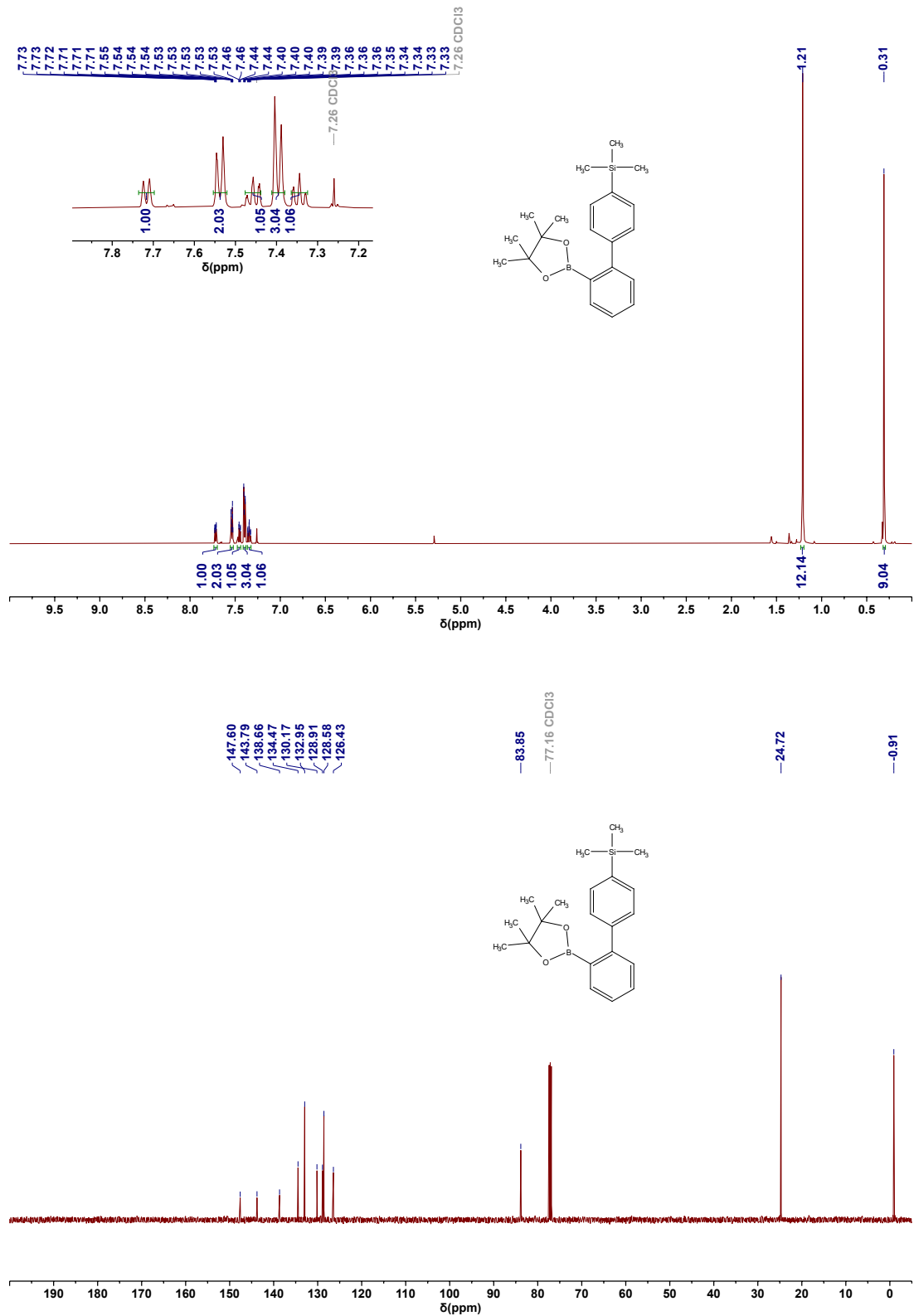

Figure S8: 1H NMR and 13C NMR spectra of compound 1c.

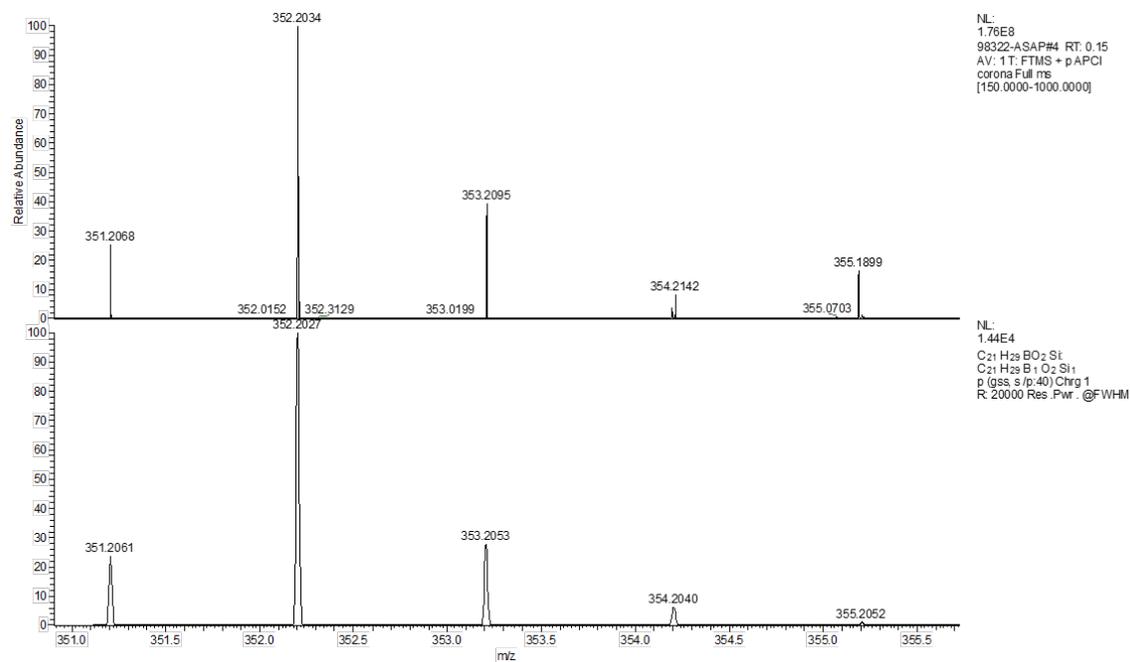

Figure S9: APCI-HRMS analysis for compound 1c.

## 2.3 Synthesis and characterization of compound 2

2',6'-dimethoxy-1,1':5',1''-ternaphthalene (**2**)

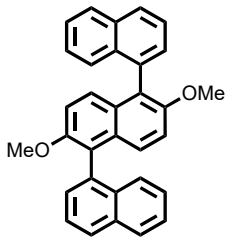

In a Schlenk tube, combine 1,5-dibromo-2,6-dimethoxynaphthalene (500 mg, 1.44 mmol), 4,4,5,5-tetramethyl-2-(naphthalen-1-yl)-1,3,2-dioxaborolane (918 mg, 3.61 mmol), $K_2CO_3$ (2 M, 3.5 mL water), toluene (20 mL), and ethanol (3.5 mL). The reaction mixture was subsequently degassed by nitrogen bubbling for 30 min at ambient temperature, after which $Pd_2(dba)_3$ (132 mg, 0.144 mmol) and SPhos (177 mg, 0.43 mmol) were added and degassed once more by nitrogen bubbling for an additional 30 min. After stirring at 110 °C for 18 h under nitrogen, the reaction mixture was allowed to cool to room temperature. The reaction mixture was poured into water, extracted with DCM, dried over $MgSO_4$, and the solvents were removed under reduced pressure. The crude product was then purified via flash column chromatography employing a hexane: DCM (5: 1) eluent, resulting in a white solid (510 mg, 1.16 mmol, yield 80%).

**APCI-HRMS**: $m/z$ 441.1851, calculated 440.1776. **$^1$H NMR** (500 MHz, $CDCl_3$) δ 8.00–7.93 (m, 4H), 7.73–7.62 (m, 2H), 7.57–7.45 (m, 5H), 7.43–7.27 (m, 5H), 7.21–7.15 (m, 2H), 3.66 (s, 6H). **$^{13}$C NMR** (126 MHz, $CDCl_3$) δ 153.18, 134.96, 133.88, 133.21, 130.01, 128.70, 128.67, 128.41, 128.39, 127.91, 127.18, 126.50, 126.45, 126.07, 125.89, 125.75, 125.74, 123.30, 123.28, 114.82, 56.99.

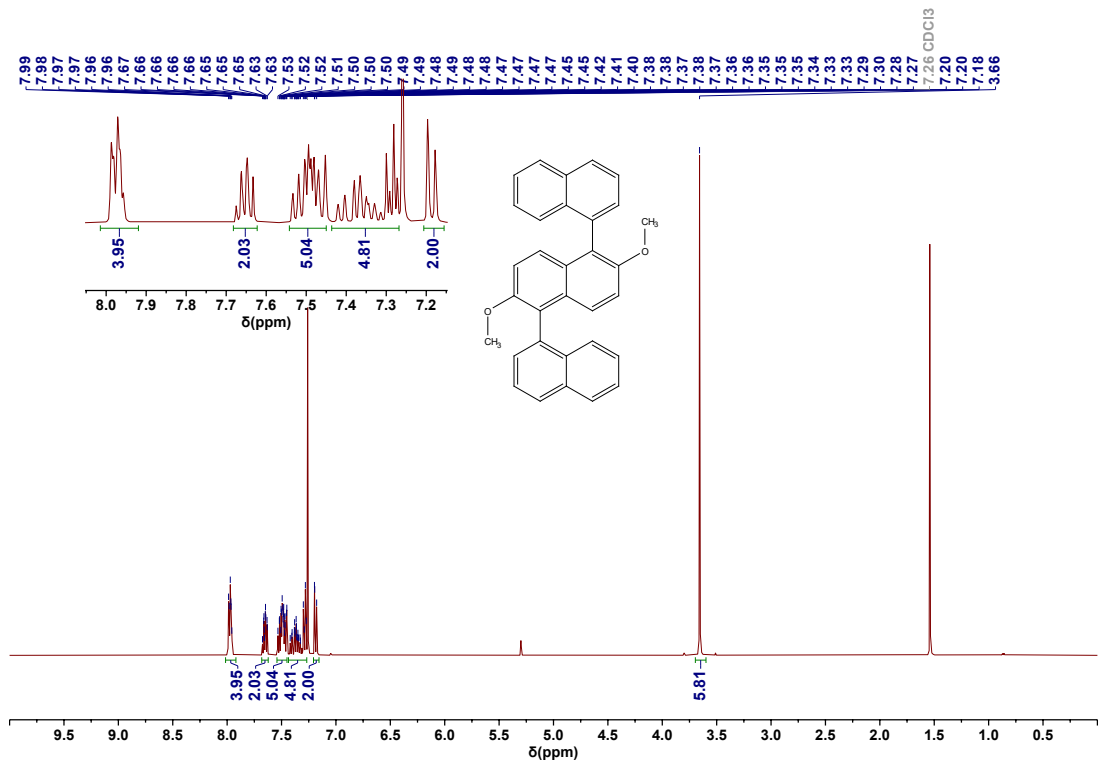

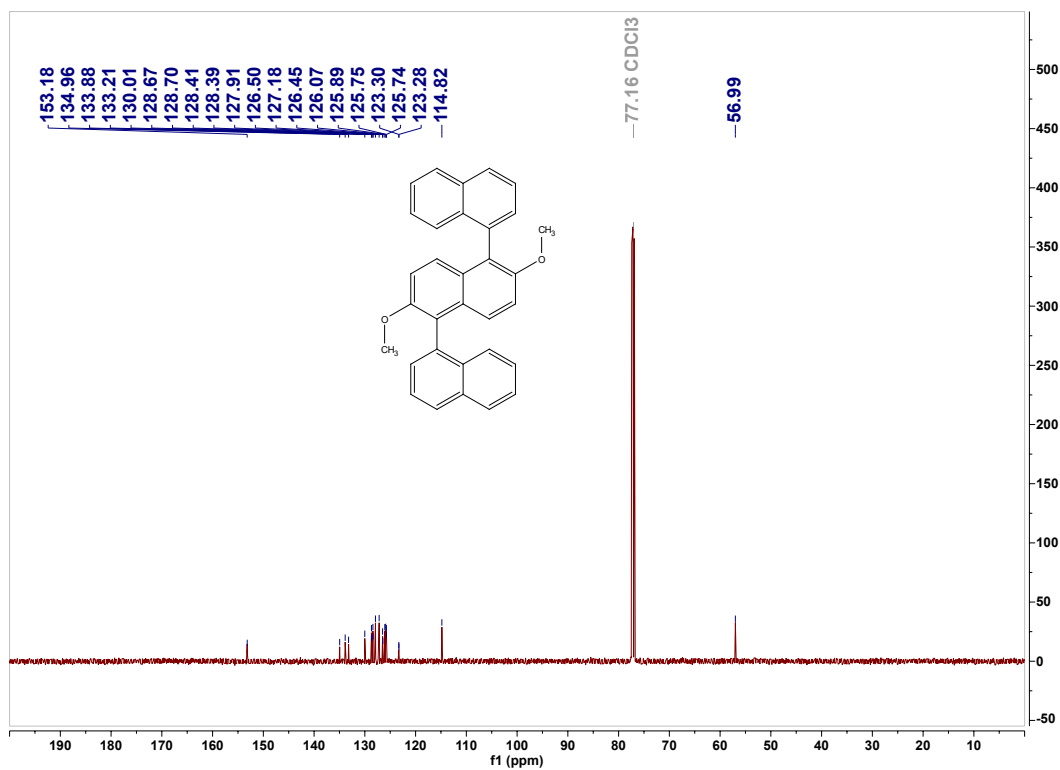

Figure S10: $^1$**H NMR** and $^{13}$**C NMR** spectra of compound **2**.

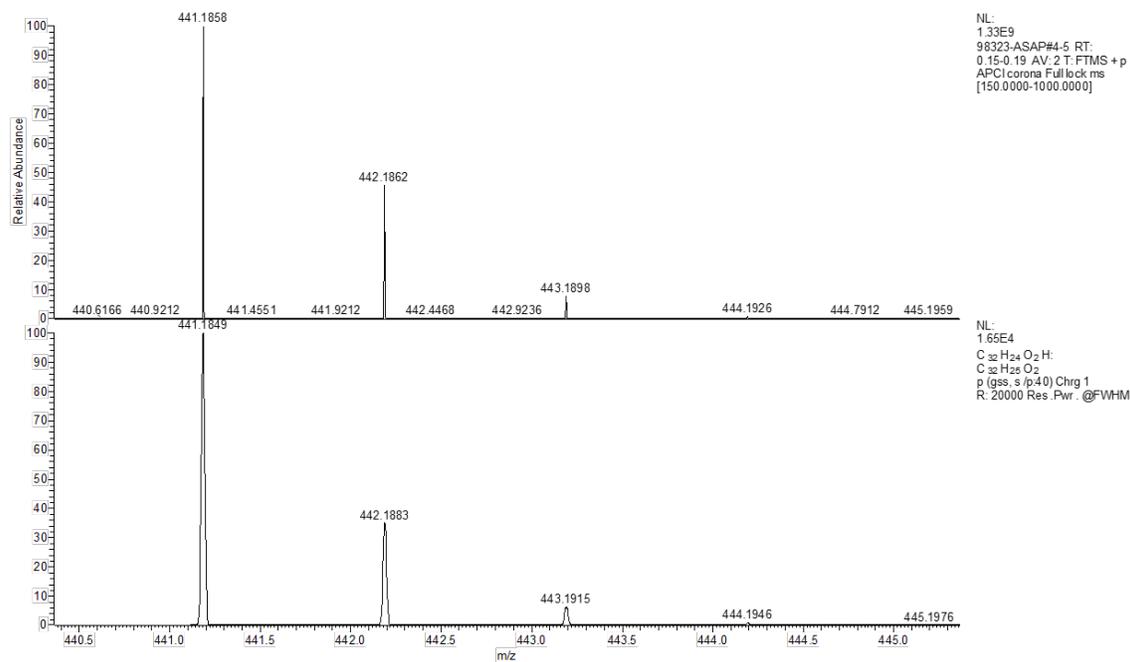

Figure S11: **APCI-HRMS** analysis for compound **2**.

## 2.4 Synthesis and characterization of compound 3'

[1,1':5',1''-ternaphthalene]-2',6'-diol (**3'**)

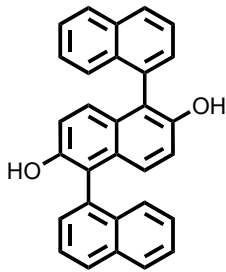

To a solution of compound **2** (500 mg, 1.13 mmol) in anhydrous DCM (90 mL) was added boron tribromide 0.65 mL at 0 °C, and then the mixture was stirred overnight at room temperature. Water was added to this mixture dropwise at 0 °C. After stirring for 15 min at room temperature, the mixture was extracted with ethyl acetate (3 × 30 mL). The organic layer was separated and dried over MgSO$_4$. The solvent was removed under vacuum, and the residue was washed with hexane to afford compound **3'** as a white solid in quantitative yield, 99% (466 mg, 1.13 mmol).

**APCI-HRMS**: $m/z$ 412.1458, calculated 412.1463. $^1$**H NMR** (500 MHz, CDCl$_3$) δ 8.06 (d, $J$ = 8.3 Hz, 2H), 8.03–7.97 (m, 2H), 7.76–7.67 (m, 2H), 7.64 (dd, $J$ = 7.0, 1.3 Hz, 1H), 7.61–7.53 (m, 2H), 7.50–7.37 (m, 3H), 7.17 (dd, $J$ = 9.1, 4.7 Hz, 2H), 7.10 (dd, $J$ = 9.1, 1.3 Hz, 2H), 4.74 (s, 2H). $^{13}$**C NMR** (126 MHz, CDCl$_3$) δ 149.35, 134.39, 133.02, 131.84, 129.44, 126.98, 126.08, 119.33.

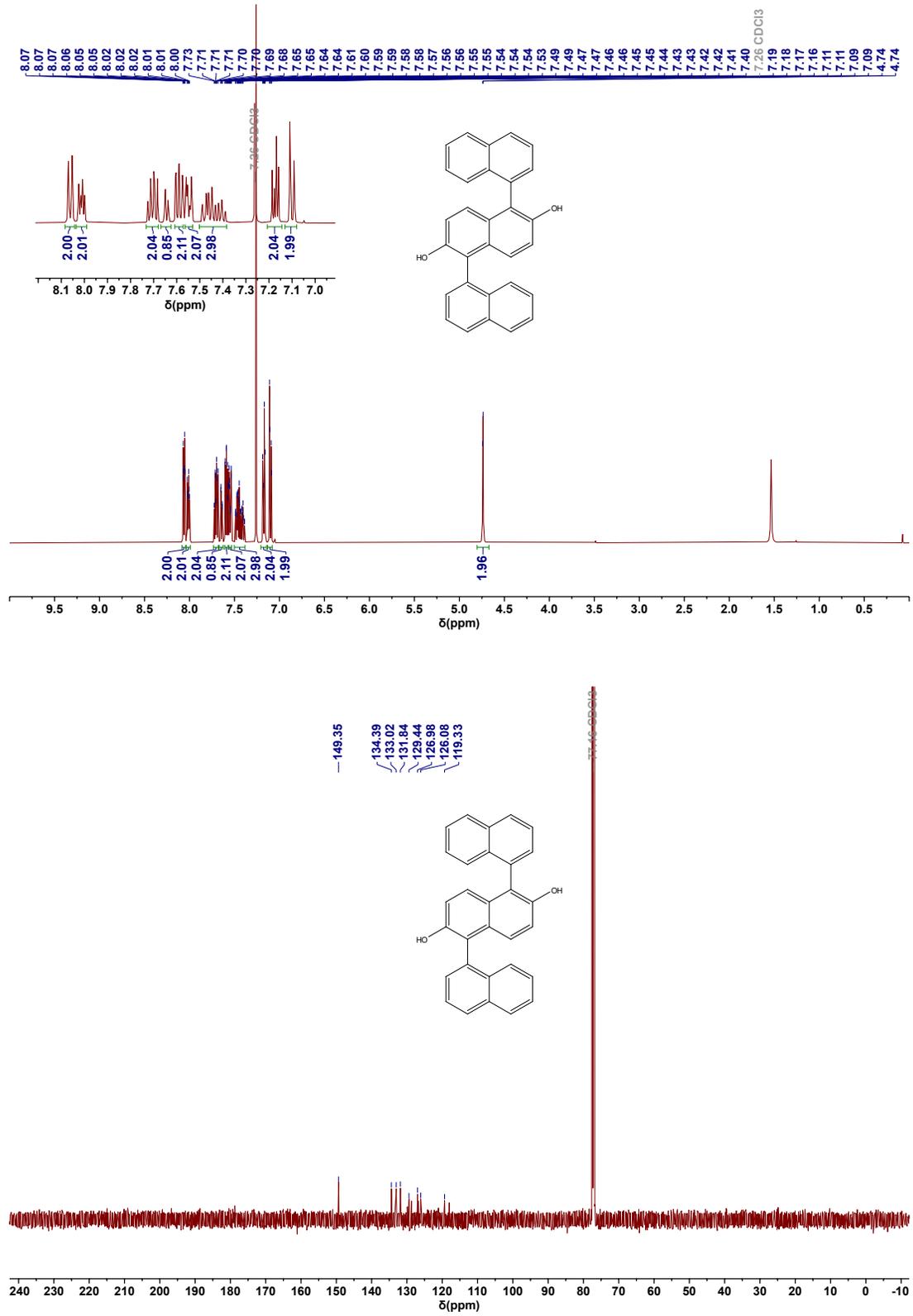

Figure S12: $^1$H NMR and $^{13}$C NMR spectra of compound **3'**.

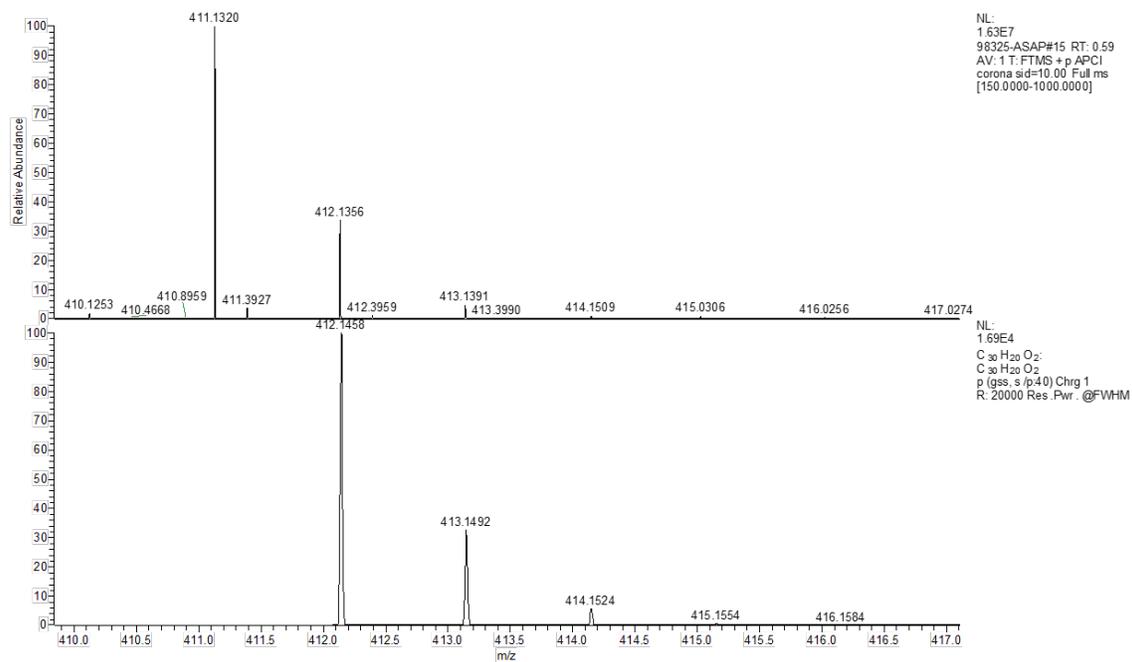

Figure S13: **APCI-HRMS** analysis for compound **3'**.

## 2.5 Synthesis and characterization of compound 3

[1,1':5',1''-ternaphthalene]-2',6'-diyl bis(trifluoromethanesulfonate) (**3**)

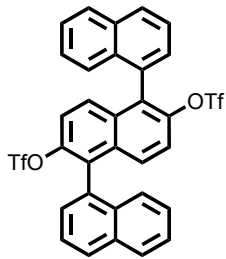

To a 250 mL round-bottom flask under nitrogen atmosphere, compound **3'** (450 mg, 1.09 mmol) and pyridine (0.35 mL, 4.36 mmol) were added and dissolved in DCM (100 mL). The reaction mixture was cooled to 0 °C and trifluoromethanesulfonic anhydride (0.73 mL, 4.36 mmol) was added. The reaction mixture was stirred at room temperature for 12 h under a nitrogen atmosphere. The reaction was quenched with water and 1 M HCl solution and extracted with DCM. The combined organic layers were washed with saturated aqueous $NaHCO_3$ solution and brine, then dried over $MgSO_4$. The solvent was removed under vacuum, affording compound **3** as a white solid (530 mg, 0.78 mmol, yield 71.86%).

**APCI-HRMS**: $m/z$ 677.0421, calculated 676.0449. **$^1$H NMR** (500 MHz, CDCl$_3$) $\delta$ 8.09 (dt, $J$ = 8.4, 1.1 Hz, 2H), 8.05–7.97 (m, 2H), 7.69 (ddd, $J$ = 8.3, 7.0, 5.8 Hz, 2H), 7.64–7.52 (m, 6H), 7.48–7.38 (m, 4H), 7.32 (dd, $J$ = 8.4, 1.1 Hz, 1H), 7.26 (s, 2H). **$^{13}$C NMR** (126 MHz, CDCl$_3$) $\delta$ 145.86, 133.81, 133.37, 133.30, 132.42, 131.51, 129.91, 129.80, 129.53, 129.40, 128.74, 128.69, 127.00, 126.52, 125.62, 125.56, 125.42, 125.40, 121.36.

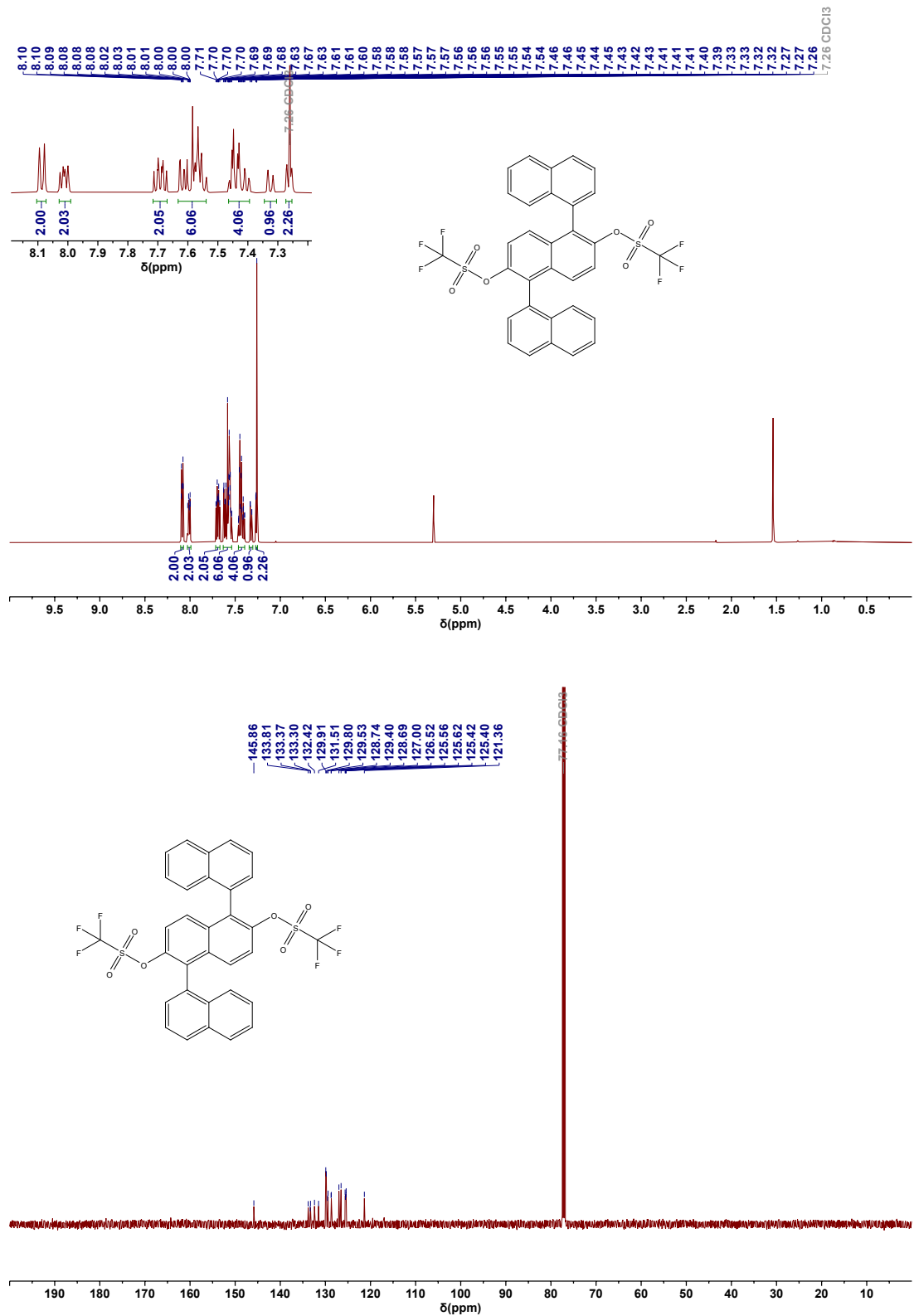

Figure S14: $^1$H NMR and $^{13}$C NMR spectra of compound **3**.

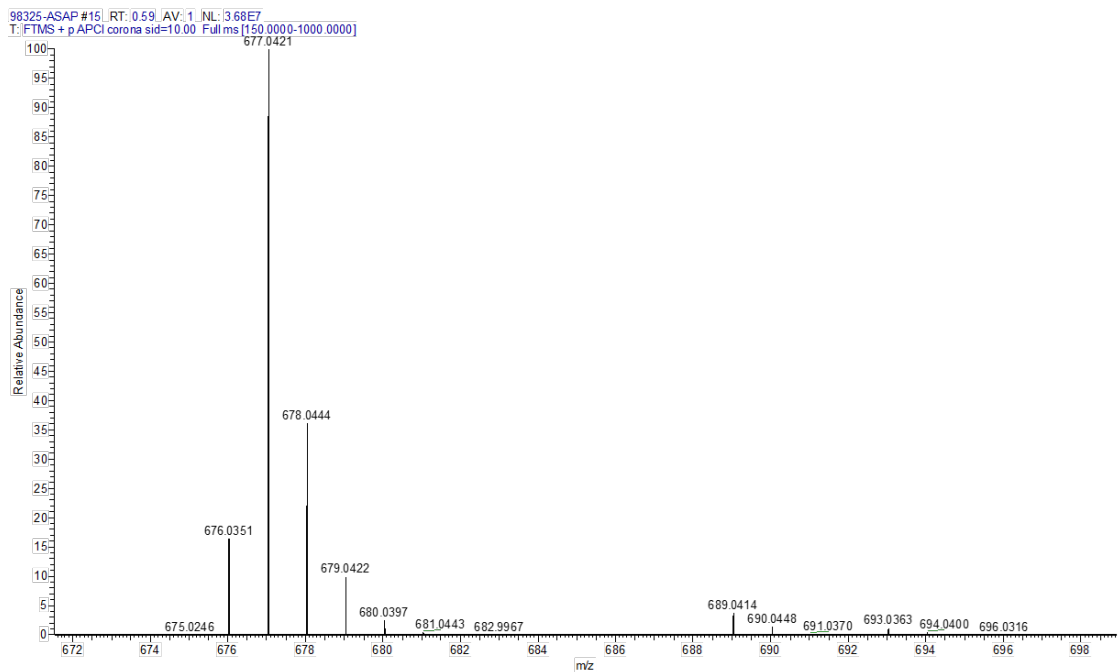

Figure S15: **APCI-HRMS** analysis for compound **3**.

## 2.6 Synthesis and characterization of compound 4

2',6'-bis(4'-(trimethylsilyl)-[1,1'-biphenyl]-2-yl)-1,1':5',1''-ternaphthalene (**4**)

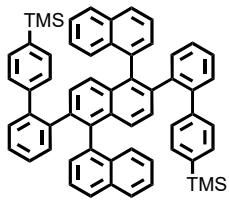

Under argon atmosphere, compound **3** (540 mg, 0.79 mmol), compound **1c** (731 mg, 0.207 mmol), and K$_3$PO$_4$ (2 g) were dissolved in dry toluene (25 mL). After degassing with nitrogen for 40 min, the catalyst Pd(dppf)Cl$_2$ (86 mg) and SPhos (48 mg) were added, followed by further degassing with nitrogen for 30 min. The mixture was heated at 100 °C for 24 h. After removal of the solvents under reduced pressure, the crude product was purified by column chromatography using flash silica gel with hexane: DCM (1: 1) as eluent to yield a white solid (520 mg, 0.68 mmol, yield 85.48%).

**APCI-HRMS**: $m/z$ 828.3602, calculated 828.3608. **$^1$H NMR** (400 MHz, CDCl$_3$) δ 7.94–7.64 (m, 4H), 7.52 (dd, $J$ = 19.8, 8.9 Hz, 2H), 7.46–7.38 (m, 3H), 7.33 (dd, $J$ = 16.8, 7.6 Hz, 5H), 7.25–6.88 (m, 15H), 6.79 (m, 2H), 6.61 (d, $J$ = 7.0 Hz, 1H), 6.45 (dd, $J$ = 32.1, 8.1 Hz, 2H), 0.34–0.16 (m, 18H). **$^{13}$C NMR** (126 MHz, CDCl$_3$) δ 142.08, 140.84, 140.28, 139.77, 138.82, 138.27, 137.50, 133.00, 132.78, 132.48, 132.09, 130.25, 130.12, 129.37, 129.20, 128.64, 127.61, 127.41, 127.21, 126.31, 126.19, 125.90, 125.52, 124.75, -0.80.

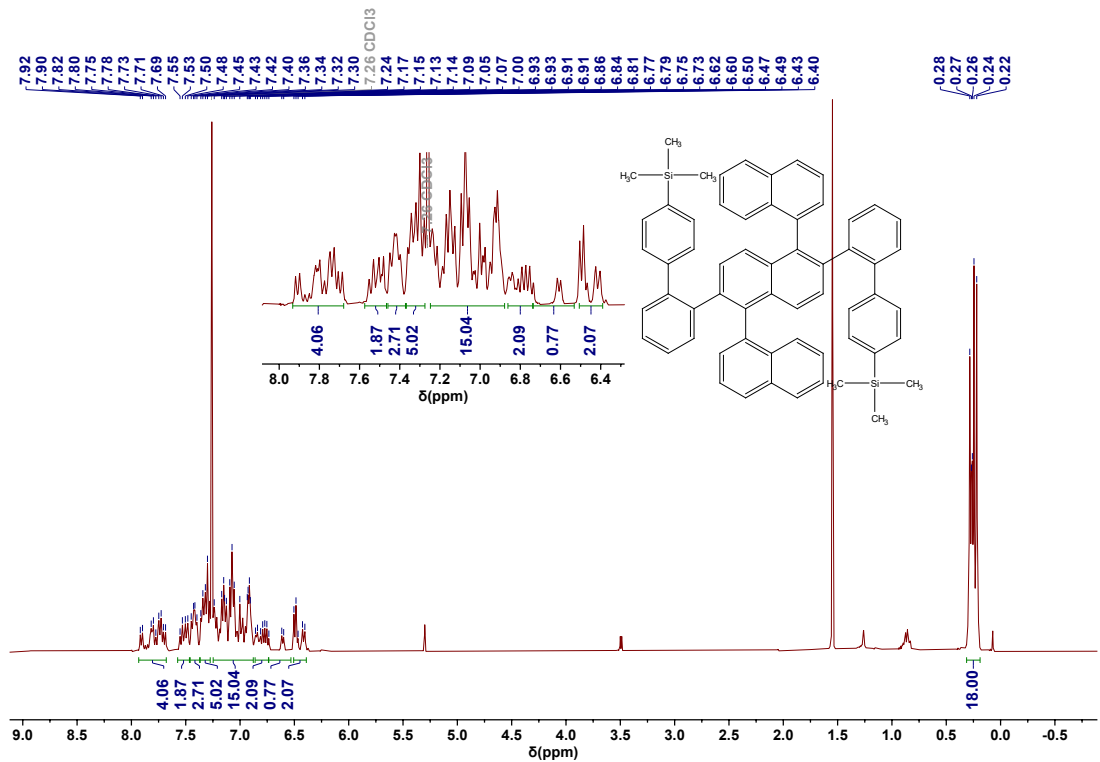
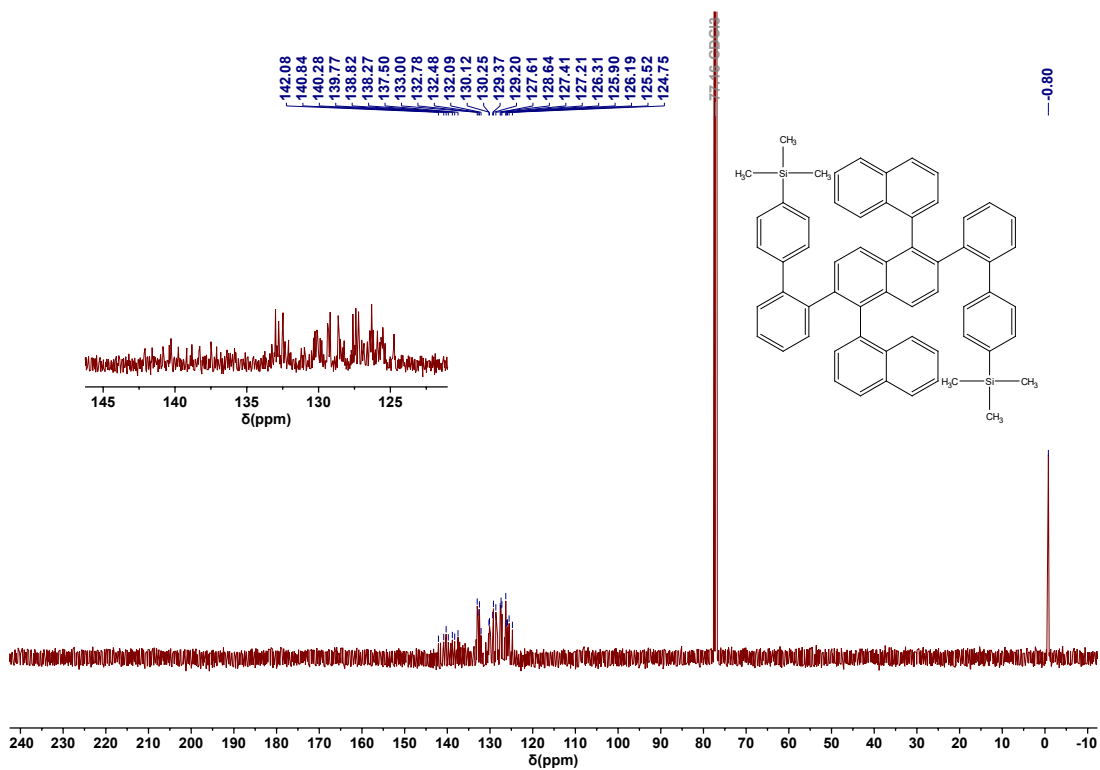

Figure S16: $^1$H NMR and $^{13}$C NMR spectra of compound **4**.

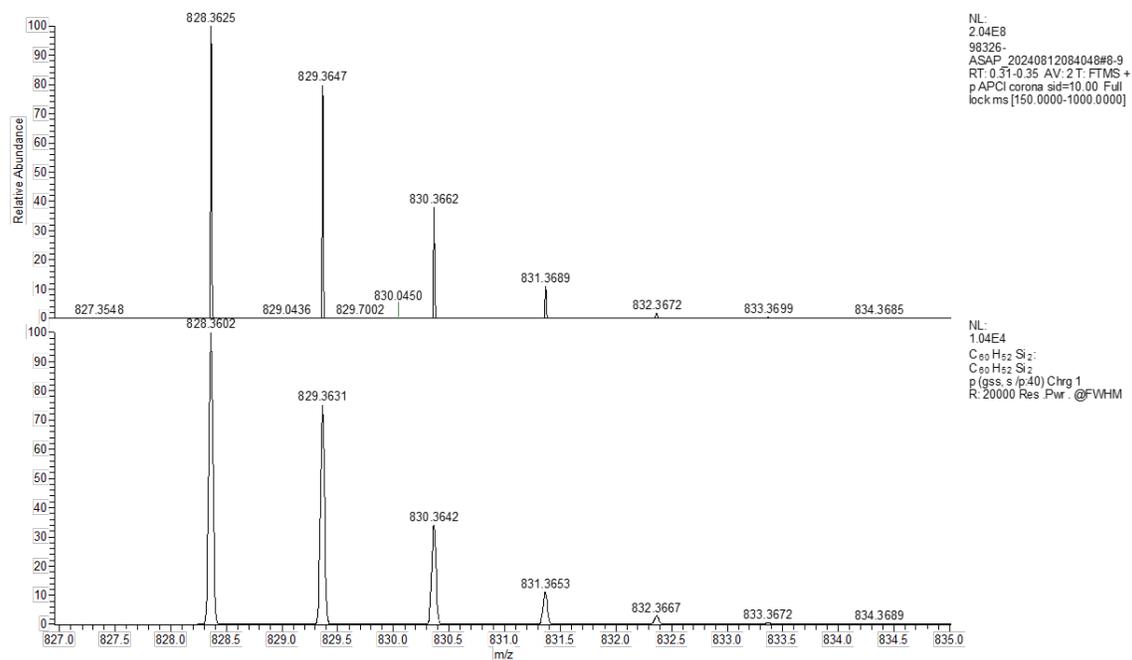

Figure S17: **APCI-HRMS** analysis for compound **4**.

## 2.7 Synthesis and characterization of compound 5

2',6'-bis(4'-iodo-[1,1'-biphenyl]-2-yl)-1,1':5',1"-ternaphthalene (**5**)

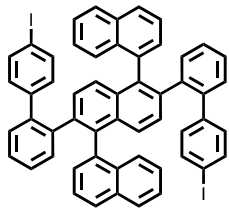

In a two-necked flask, compound **1** (1.0 eq, 150 mg, 0.180 mmol) was dissolved in 150 mL of anhydrous DCM under a nitrogen atmosphere. The solution was subsequently cooled to $-70\,^\circ$C, after which ICl solution (2.7 eq, 76 mg, 0.47 mmol) was added. The reaction mixture was stirred for an additional 2 h and allowed to reach room temperature. The reaction was subsequently quenched using a saturated aqueous solution of $Na_2S_2O_3$, followed by extraction with DCM. After removal of the solvents under reduced pressure, the crude product was purified by column chromatography using flash silica gel with DCM as eluent to yield a white solid (156 mg, 0.16 mmol, yield 92.52%).

**APCI-HRMS**: $m/z$ 936.0744, calculated 936.0750. $^1$**H NMR** (500 MHz, CDCl$_3$) $\delta$ 7.91–7.67 (m, 4H), 7.58–7.41 (m, 6H), 7.40–7.28 (m, 4H), 7.25–7.13 (m, 4H), 7.12–7.01 (m, 6H), 6.98–6.82 (m, 4H), 6.81–6.57 (m, 3H), 6.41 (t, $J = 8.8$ Hz, 1H), 6.12 (m, 2H). $^{13}$**C NMR** (126 MHz, CDCl$_3$) $\delta$ 145.86, 133.84, 133.39, 132.41, 131.53, 130.33, 129.97, 129.93, 129.82, 129.55, 128.76, 127.04, 126.56, 125.58, 125.44, 125.38, 121.38, 119.66.

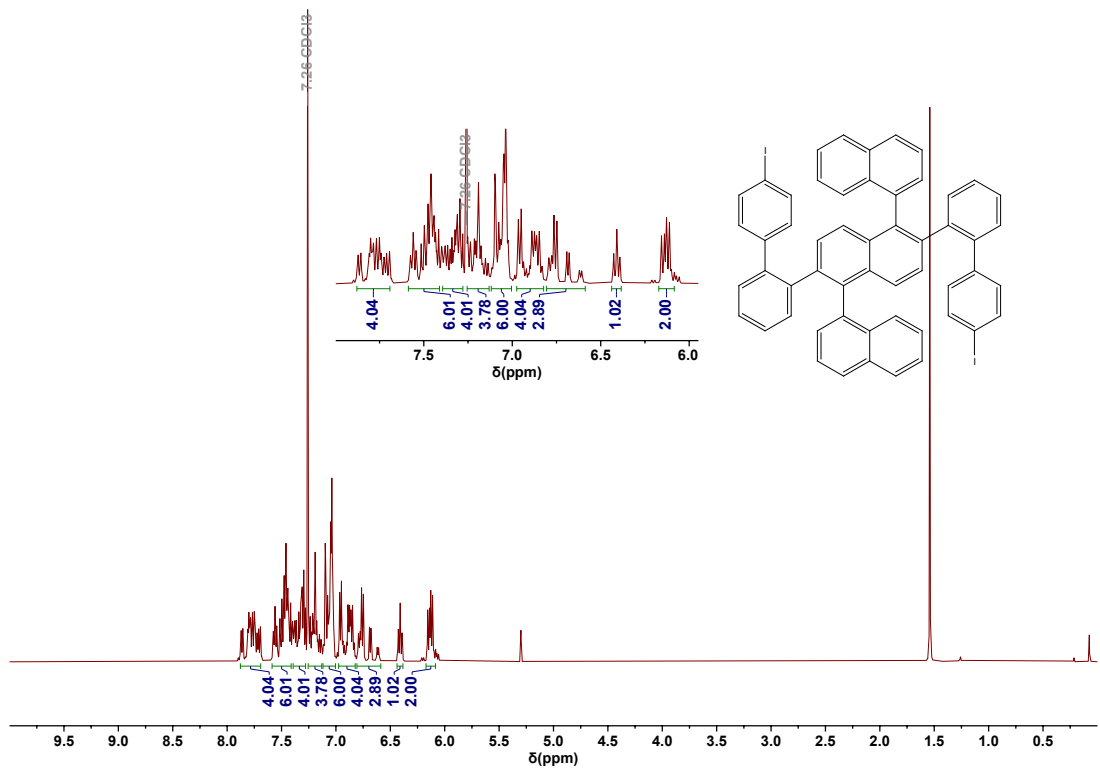
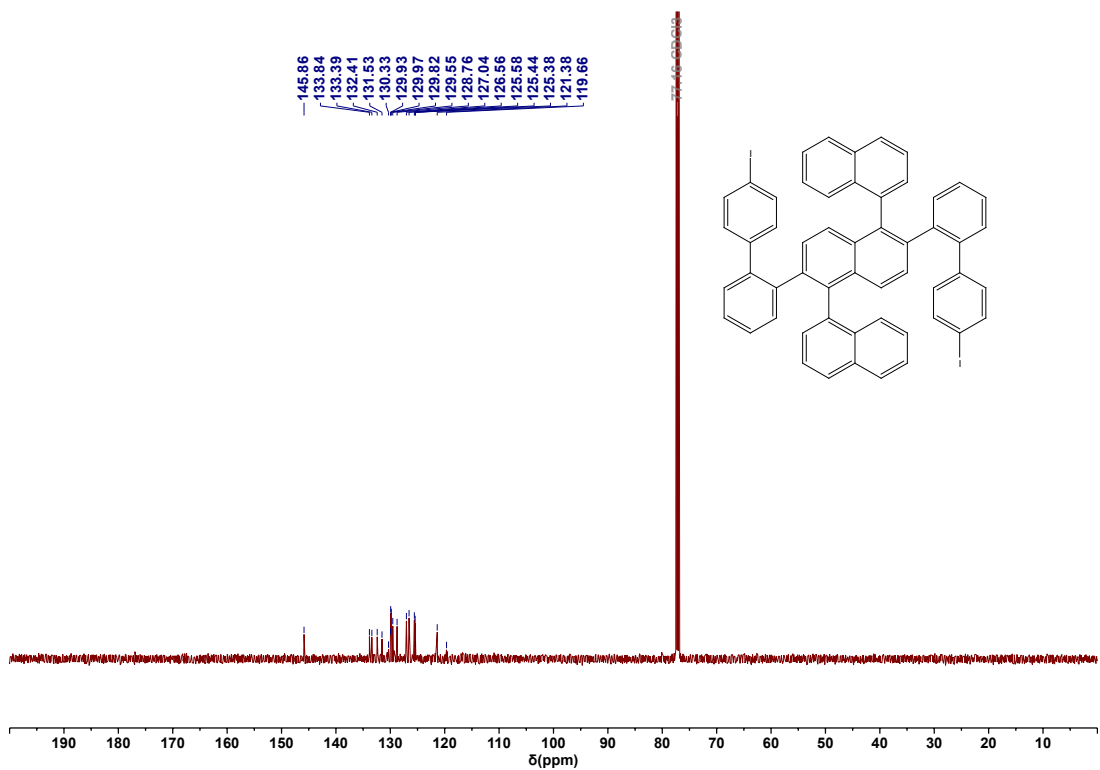

Figure S18: $^1$**H NMR** and $^{13}$**C NMR** spectra of compound **5**.

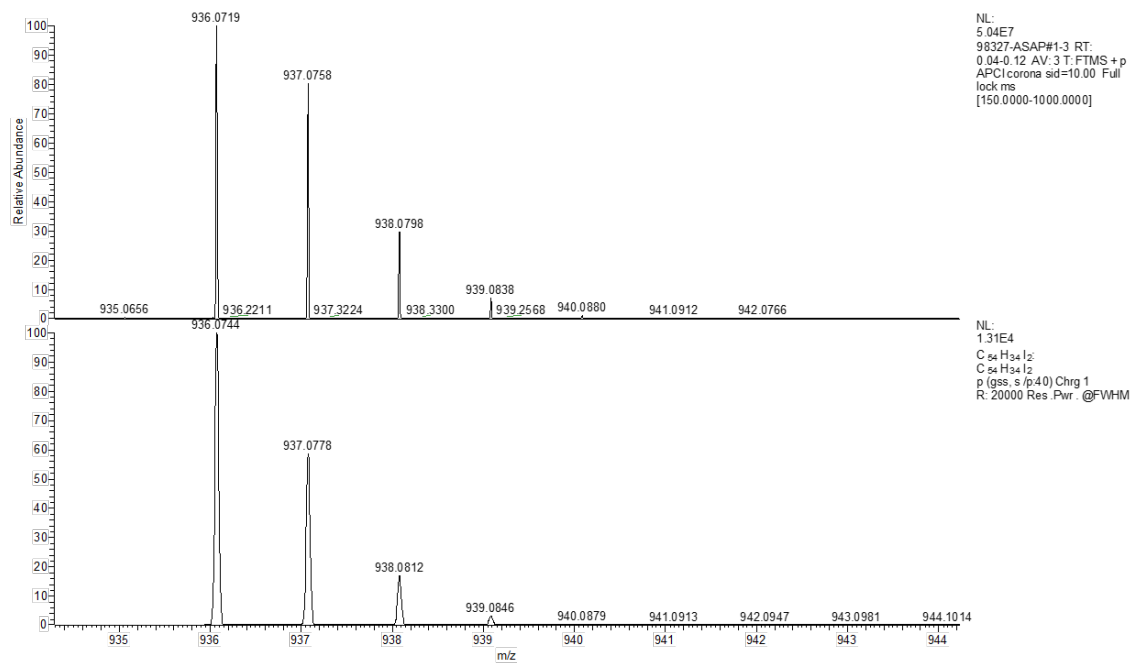

Figure S19: **APCI-HRMS** analysis for compound **5**.

## 2.8 X-ray diffraction data and crystal structure determination of compound 5

X-ray diffraction data for compound **5** were collected using beamline i19 EH1 at Diamond Light Source using Zr L-edge ($\lambda = 0.6889$ Å) radiation, equipped with a 3-circle Dual Air-Bearing Fixed-$\chi$ goniometer, a Pilatus 2M detector, and an Oxford Cryosystems 800 Plus nitrogen gas-flow system, at a temperature of 100 K. Data for 5 were collected using Generic Data Acquisition (GDA) [10, 11].

Data were reduced using CRYSALISPRO v42 [12]. Absorption correction was performed using empirical methods (SCALE3 ABSPACK) based on symmetry-equivalent reflections combined with measurements at different azimuthal angles.

The crystal structure was solved and refined against all $F^2$ values using the SHELX and OLEX2 suite of programs [13–15]. Coordinates and anisotropic displacement parameters for all non-hydrogen atoms were freely refined. Hydrogen atoms were constrained to idealised positions with coordinates refined to ride on the parent atom. Hydrogen isotropic atomic displacement parameters were constrained to ride on the parent atom with an appropriate multiplier for the hybridisation.

CCDC 2537291 contains the supplementary crystallographic data for this paper. These data can be obtained free of charge via https://www.ccdc.cam.ac.uk/structures/, or by emailing data_request@ccdc.cam.ac.uk, or by contacting The Cambridge Crystallographic Data Centre, 12 Union Road, Cambridge, CB2 1EZ, UK; fax: +44 1223 336033.

| Parameter | Value | | |
|---|---|---|---|
| Identification code | m_Compound 5_c1_c1_ | | |
| Empirical formula | $C_{54}H_{34}I_2$ | | |
| Formula weight | 936.61 | | |
| Temperature / K | 100.00(10) | | |
| Crystal system | monoclinic | | |
| Space group | $P2_1/n$ | | |
| $a$ / Å | 15.7695(4) | | |
| $b$ / Å | 8.8118(2) | | |
| $c$ / Å | 28.3891(7) | | |
| $\alpha$ / ° | 90 | | |
| $\beta$ / ° | 94.796(2) | | |
| $\gamma$ / ° | 90 | | |
| Volume / Å$^3$ | 3931.05(17) | | |
| Z | 4 | | |
| $\rho_{calc}$ / g cm$^{-3}$ | 1.583 | | |
| $\mu$ / mm$^{-1}$ | 1.519 | | |
| F(000) | 1856.0 | | |
| Crystal size / mm$^3$ | 0.124 × 0.123 × 0.015 | | |
| Radiation | synchrotron ($\lambda$ = 0.6889 Å) | | |
| 2Θ range / ° | 2.974 to 51.002 | | |
| Index ranges | $-19 \leq h \leq 16$ | $-10 \leq k \leq 11$ | $-35 \leq l \leq 35$ |
| Reflections collected | 34986 | | |
| Independent reflections | 7940 [$R_{int}$ = 0.0820, $R_\sigma$ = 0.0688] | | |
| Data/restraints/parameters | 7940 / 0 / 505 | | |
| Goodness-of-fit on $F^2$ | 1.060 | | |
| Final R indexes [I ≥ 2σ(I)] | $R_1$ = 0.0686 | $wR_2$ = 0.1832 | |
| Final R indexes [all data] | $R_1$ = 0.1025 | $wR_2$ = 0.2003 | |
| Largest diff. peak / hole / e Å$^{-3}$ | 2.10 | -1.07 | |

Table S1: Crystallographic information for compound **5**.

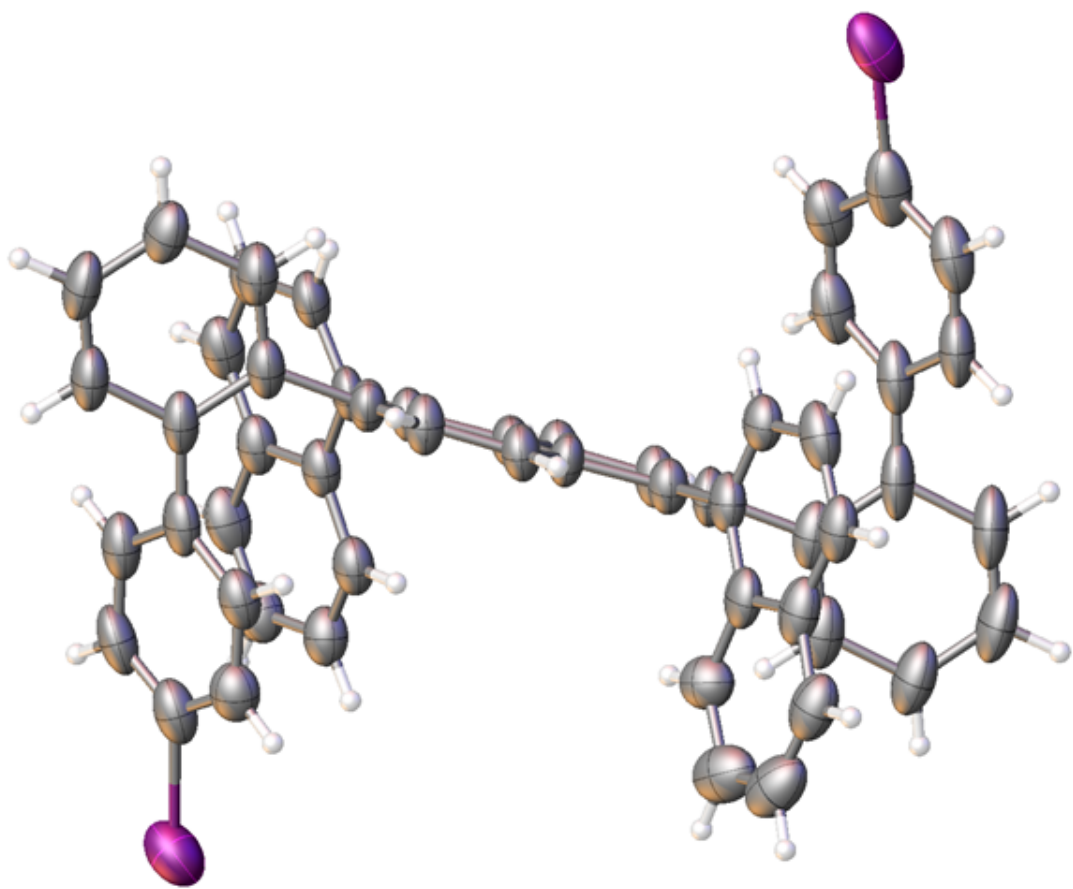

Figure S20: Structure of compound **5** in the crystal. Grey = carbon, white = hydrogen, purple = iodine. Thermal ellipsoid probability is 50%.

# 3 Electronic structure characterization of (4,2,7)-chGNR

## 3.1 Electronic characterization of VB–1 and CB+1

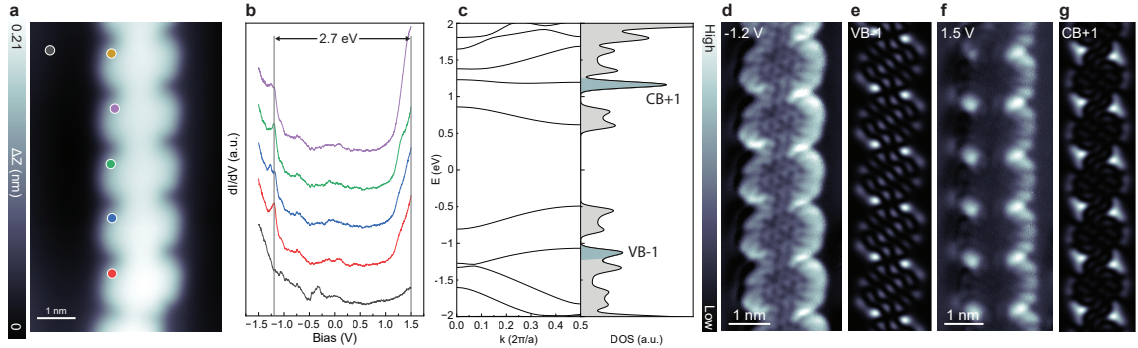

Figure S21: Electronic characterization of the VB-1 and CB+1 of (4,2,7)-chGNR. **(a,b)** The STS measurements in (b) were acquired on the 5-unit (4,2,7)-chGNR segment shown in (a). The spectra are color-coded to their STS positions in (a). A CO-functionalized tip was used to acquire the dI/dV maps. **(c)** Band structure (energy versus wave vector k) (left, black; a is the lattice parameter) and density of states (right, grey) calculated for an infinitely long (4,2,7)-chGNR with Kohn–Sham DFT. Valence band - 1 (VB-1) and conduction band + 1 (CB+1) are highlighted. **(d,f)** Constant current dI/dV maps acquired over a 5-unit (4,2,7)-chGNR segment at the energy of VB-1 and CB+1. STS spectras and dI/dV maps were acquired on two different ribbons. A CO-functionalized tip is used to acquire the dI/dV maps. **(e,g)** Simulated PDOS maps within the valence and conduction bands for an infinitely long (4,2,7)-chGNR, calculated at a tip–sample distance of 3.203 Å.

## 3.2 Constant height dI/dV maps

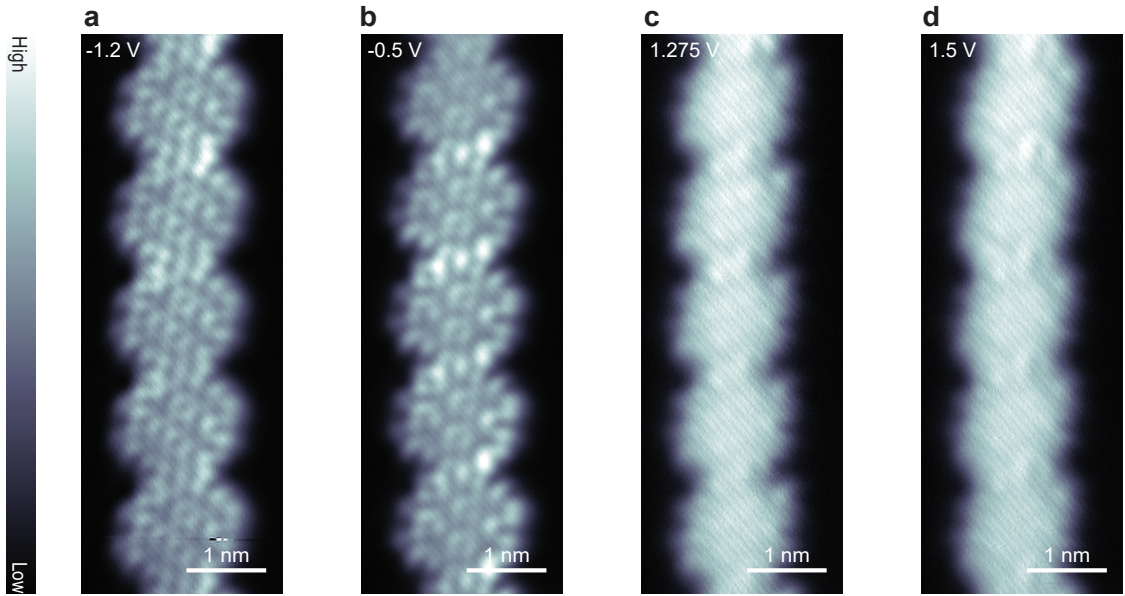

Figure S22: Constant height dI/dV maps acquired over a 5-unit (4,2,7)-chGNR segment at the energy of the VB-1, VB, CB and CB+1 (a-d, respectively). A CO-functionalized tungsten tip was used to acquire the dI/dV maps. Experimental details are provided in the Methods section in the main text.

# 4 Large-scale STM image of a high-coverage (4,2,7)-chGNR sample on an Au (111) substrate

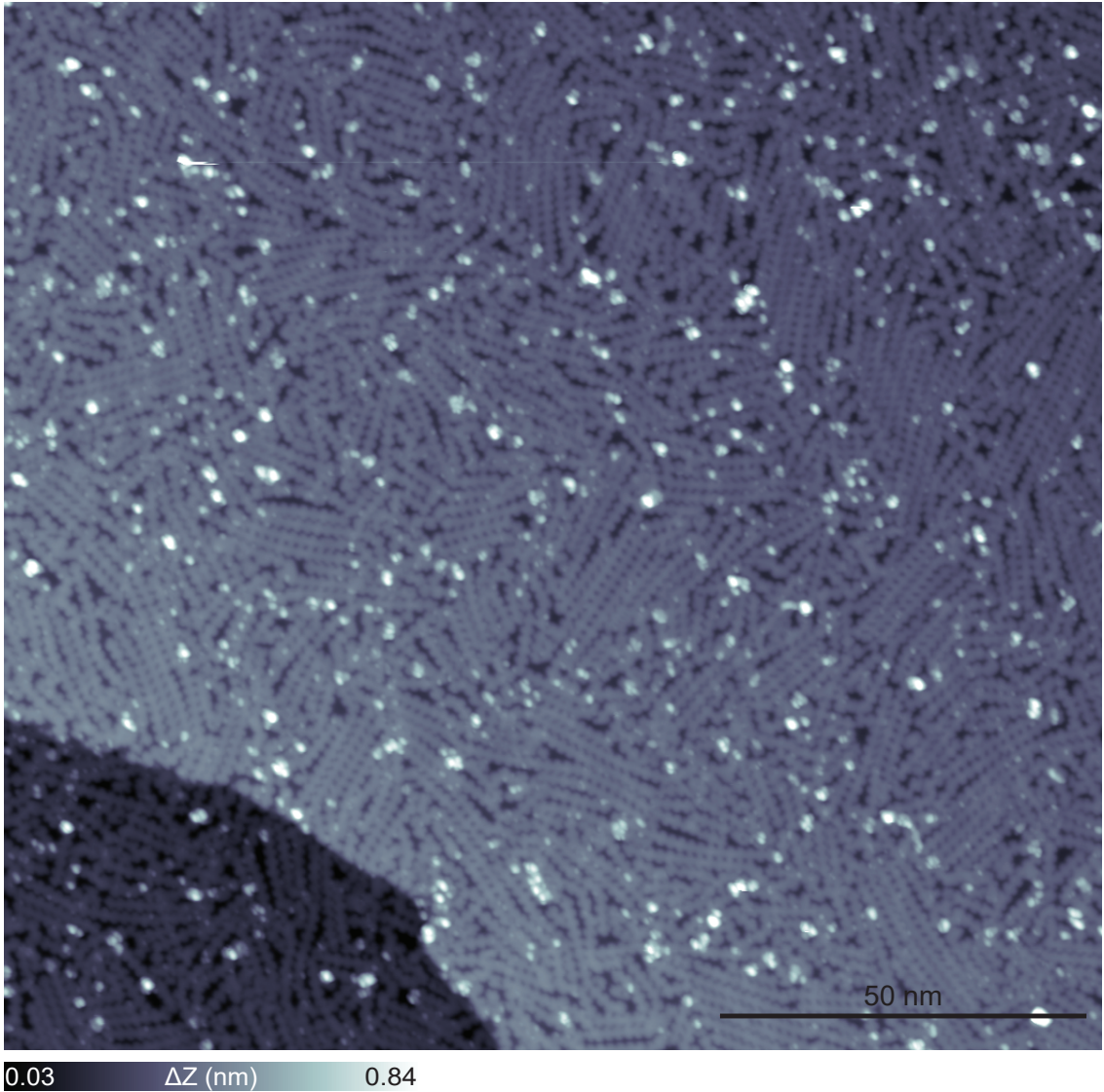

Figure S23: A large-scale STM topography image of a high-coverage (4,2,7)-chGNR sample on an Au(111) substrate. This sample was used for Raman characterization. The STM image was obtained at $4.5\,\mathrm{K}$. Scanning parameters: $-1\,\mathrm{V}$, $10\,\mathrm{pA}$. The image was processed by applying a mean-plane subtraction to the topography data.

# 5 Raman spectroscopy of (4,2,7)-chGNR

## 5.1 Simulated Raman response

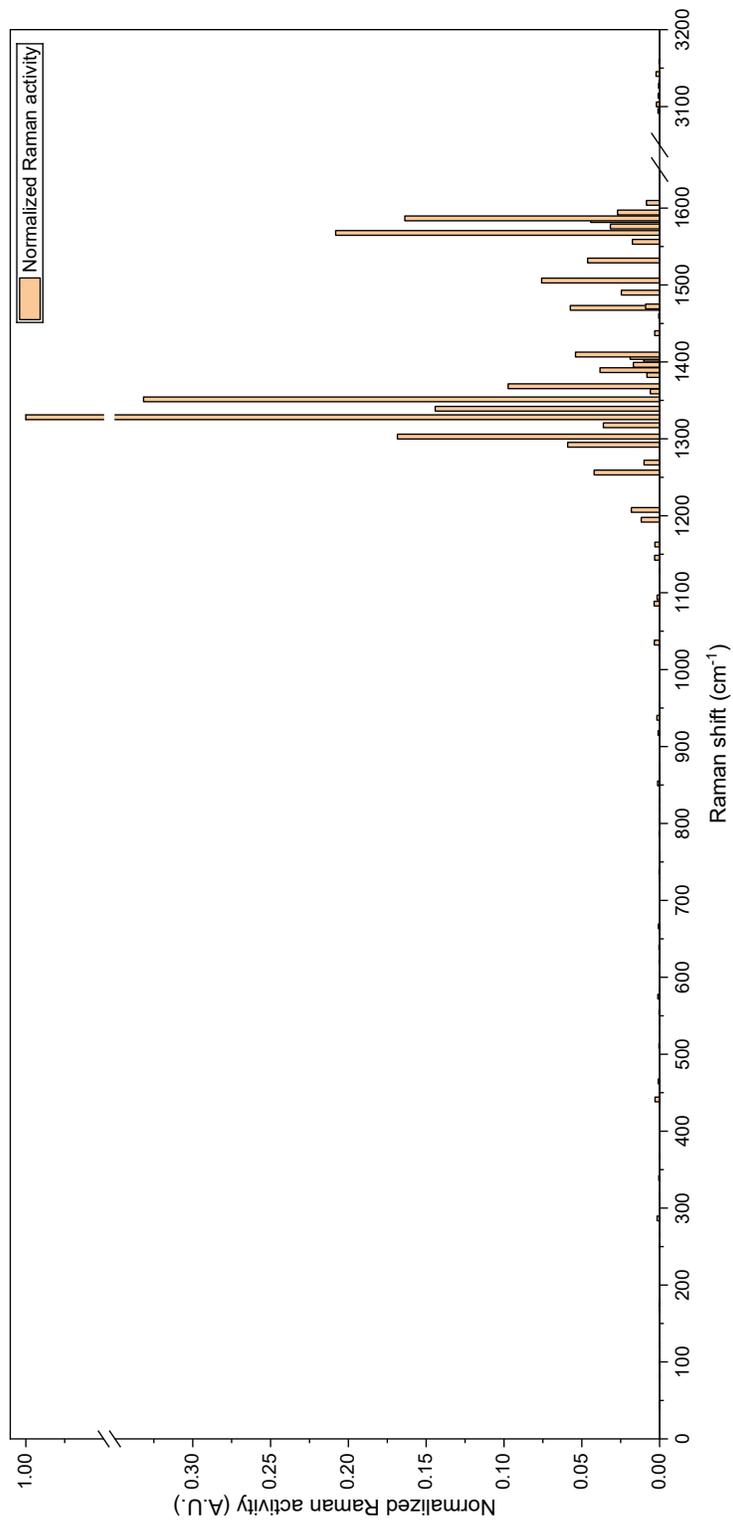

Figure S24: Simulated Raman intensity spectrum

34

## 5.2 Experimental Raman spectra

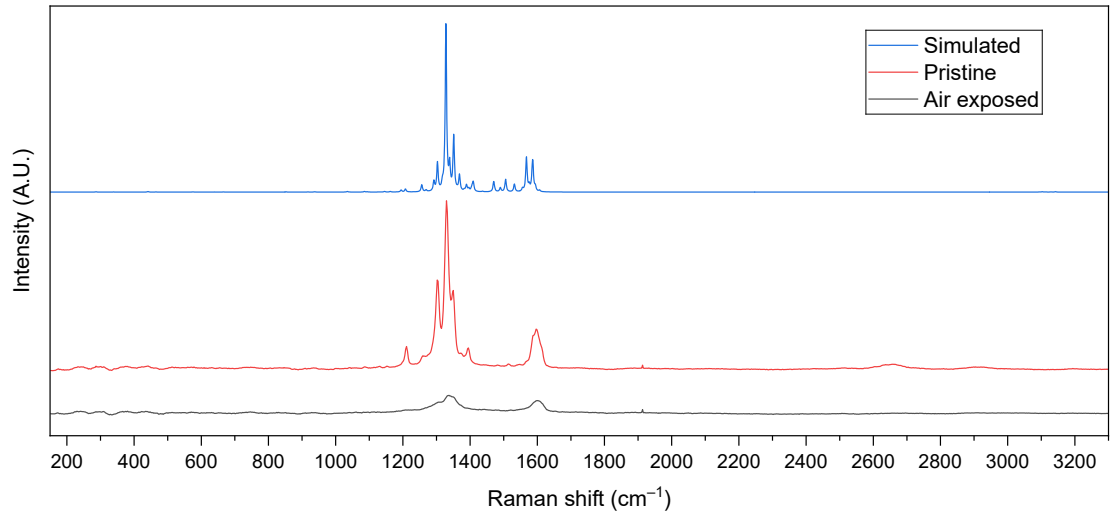

Figure S25: Full Raman spectrum of (4,2,7)-chGNR acquired under 532 nm excitation, spanning from $300\,\mathrm{cm}^{-1}$ to $3300\,\mathrm{cm}^{-1}$. The plot compares experimental spectra of pristine and air-exposed gchGNR with the DFT-simulated Raman spectrum. The simulated spectrum was generated by summing Lorentzian peaks with a broadening of $5\,\mathrm{cm}^{-1}$ at the calculated vibrational frequencies. For each spectrum, Raman mapping was performed by recording 400 individual spectra (1 s integration time) over a 20 × 20 grid (20 µm × 20 µm). The average of these spectra was used for analysis. A 600 g/mm grating was employed. All other experimental conditions were identical to those described in the Methods section of the main text. Details of the DFT simulations used to obtain the Raman spectra are also provided in the Methods section.

## 5.3 Mode assignment table

| Band | Experimental modes (cm$^{-1}$) | Simulated modes (cm$^{-1}$) | Normalized simulated intensity |
|---|---|---|---|
| | 1128 | 1129 | 2.55E−04 |
| | 1153 | 1145 | 3.19E−03 |
| | | 1163 | 3.00E−03 |
| | | 1183 | 1.60E−04 |
| | 1196 | 1195 | 1.18E−02 |
| | **1210** | **1208** | 1.82E−02 |
| | 1258 | 1256 | 4.21E−02 |
| | 1273 | 1269 | 1.00E−02 |
| | | 1292 | 5.91E−02 |
| | 1302 | 1303 | 1.69E−01 |
| | | 1318 | 3.62E−02 |
| D band | **1330** | **1328** | 1.00E+00 |
| | | 1339 | 1.44E−01 |
| | 1349 | 1351 | 3.32E−01 |
| | | 1362 | 6.02E−03 |
| | 1375 | 1368 | 9.73E−02 |
| | | 1383 | 8.17E−03 |
| | 1394 | 1389 | 3.83E−02 |
| | | 1396 | 1.69E−02 |
| | | 1404 | 1.03E−02 |
| | | 1406 | 1.89E−02 |
| | | 1410 | 5.41E−02 |
| | | 1437 | 3.13E−03 |
| | | 1460 | 5.76E−04 |
| | 1483 | 1470 | 5.74E−02 |
| | | 1472 | 9.01E−03 |
| | | 1490 | 2.46E−02 |
| | 1515 | 1506 | 7.58E−02 |
| | | 1532 | 4.63E−02 |
| | 1569 | 1556 | 1.75E−02 |
| | **1586** | **1568** | 2.08E−01 |
| | 1594 | 1576 | 3.17E−02 |
| G band | 1599 | 1584 | 4.44E−02 |
| | 1606 | 1587 | 1.64E−01 |
| | 1613 | 1594 | 2.70E−02 |
| | | 1607 | 8.44E−03 |

Table S2: Experimental and simulated vibrational modes of (4,2,7)-chGNR in the high-frequency region. Due to the limited resolution of the experimental measurements, some modes cannot be individually resolved, and the assignment of experimental modes to their simulated counterparts may therefore contain some uncertainty. The modes whose visualizations are shown in the main text are highlighted in bold. Visualizations of all other modes in the D band and G band are provided in Figures S26 and S27.

# 6 Vibrational modes visualization

## 6.1 CH and D-band modes of (4,2,7)-chGNR

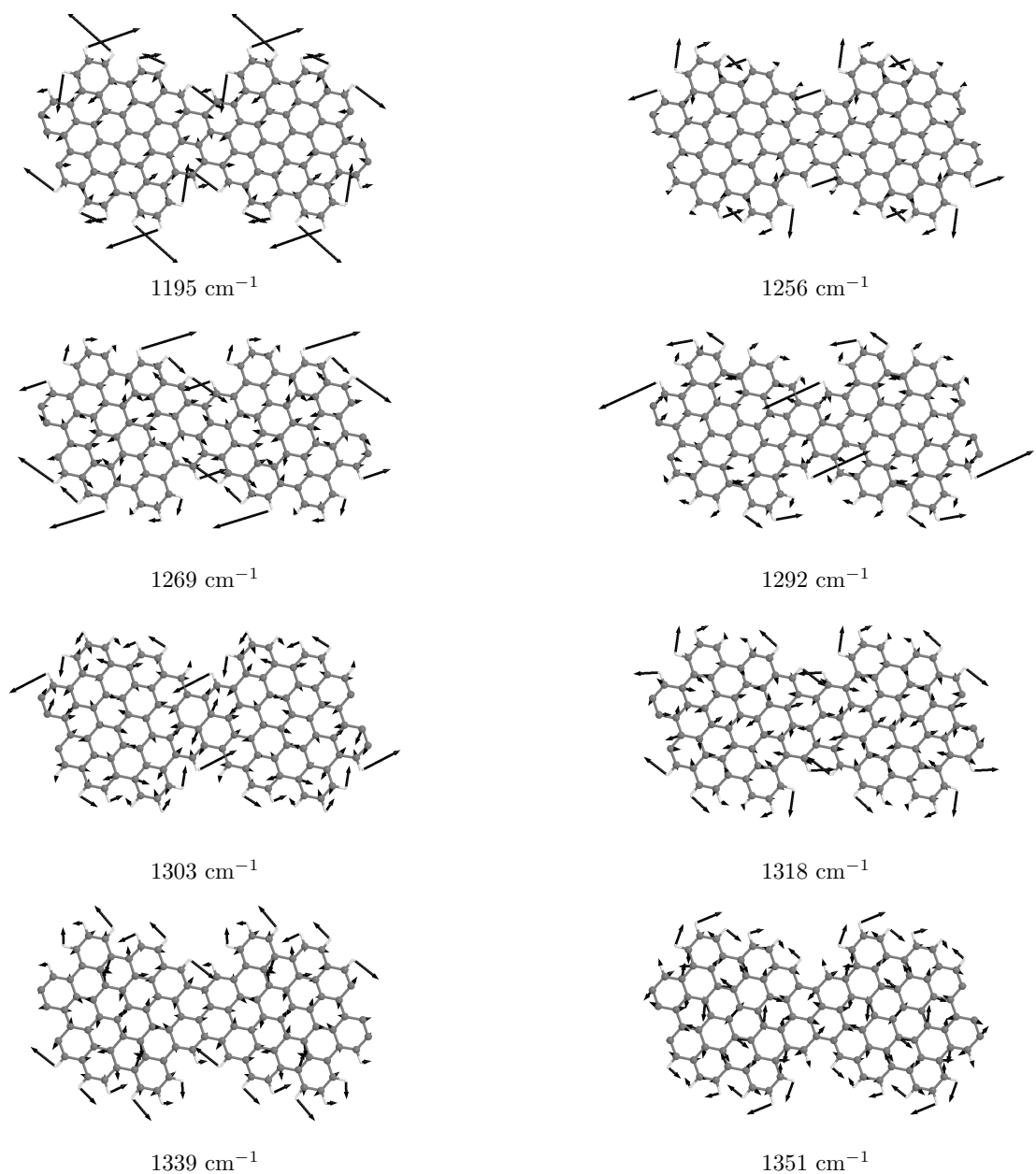

Figure S26: Vibrational modes of (4,2,7)-chGNR in the CH/D band (simulated wavenumbers). The simulated wavenumbers are shown below each panel.

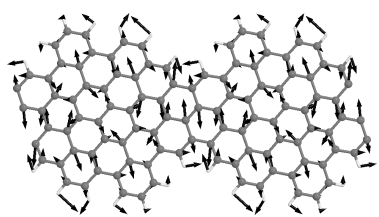

1368 cm$^{-1}$

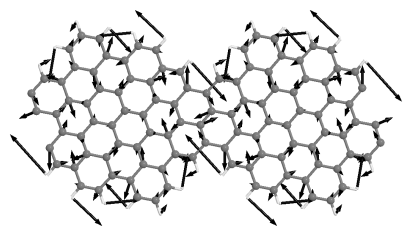

1383 cm$^{-1}$

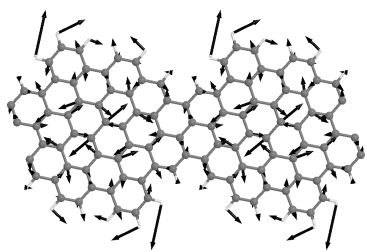

1389 cm$^{-1}$

Figure S26: (continued)

## 6.2 G-band modes

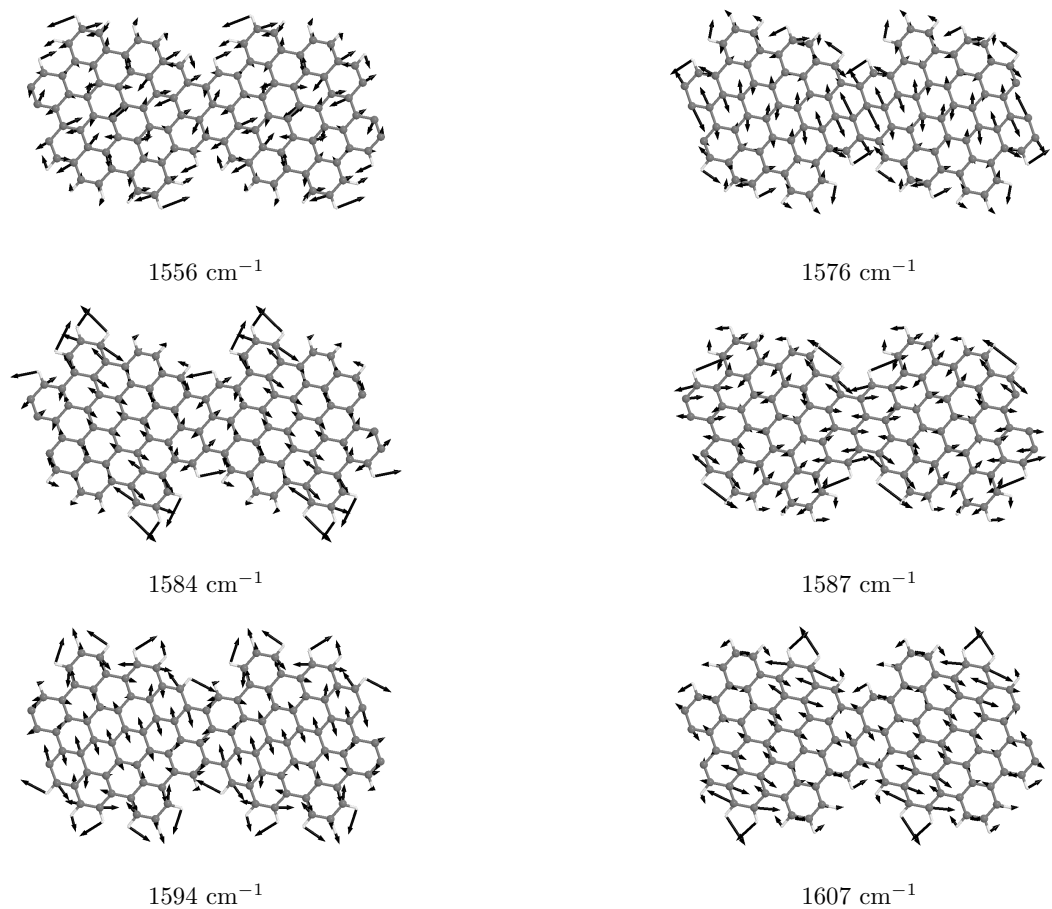

Figure S27: Vibrational modes of (4,2,7)-chGNR in the G band (simulated wavenumbers). The simulated wavenumbers are shown below each panel.

39



\end{document}